\pdfoutput=1
\documentclass[a4paper,12pt,margin=in]{article}
\usepackage{lineno}
\usepackage{natbib}
\usepackage[colorlinks,citecolor=blue,urlcolor=blue]{hyperref}
\usepackage{epsf}
\usepackage{epsfig}
\usepackage{multirow}
\usepackage[english]{babel}
\usepackage{graphicx}
\usepackage{geometry}
\usepackage{subcaption}
\usepackage{natbib}
\usepackage{tikz}
\usetikzlibrary{positioning, arrows.meta,fit}
\usepackage{tabularx}
\usepackage{amssymb}
\usepackage{amsmath}
\usepackage{color}
\usepackage{bm}
\usepackage{enumerate}
\usepackage{authblk}
\usepackage{float}
\usepackage{geometry}
 \usepackage{setspace}
\usepackage[font=footnotesize,labelfont=bf]{caption}

\newcommand{\specificthanks}[1]{\@fnsymbol{#1}}

\title{Integrated differential analysis of multi-omics data using a joint mixture model: idiffomix}
\author[1]{\normalsize{Koyel Majumdar}}
\author[2]{Florence Jaffr\'ezic}
\author[2]{Andrea Rau}
\author[1]{Isobel Claire Gormley\thanks{claire.gormley@ucd.ie}}
\author[1]{Thomas Brendan Murphy}
\affil[1]{\small{School of Mathematics and Statistics,University College Dublin, Ireland.}}
\affil[2]{INRAE, UMR1313 AgroParisTech, GABI, Universit Paris-Saclay, France.}

\date{}

\begin{document}
\nolinenumbers
	\doublespacing
\maketitle

\begin{abstract}
Gene expression and DNA methylation are two interconnected biological processes and understanding their relationship is important in advancing understanding in diverse areas, including disease pathogenesis, environmental adaptation, developmental biology, and therapeutic responses. Differential analysis, including the identification of differentially methylated cytosine-guanine dinucleotide (CpG) sites (DMCs) and differentially expressed genes (DEGs) between two conditions, such as healthy and affected samples, can aid understanding of biological processes and disease progression. Typically, gene expression and DNA methylation data 
are analysed independently to identify DMCs and DEGs which are further analysed to explore relationships between them, or methylation data are aggregated to gene level for analysis. Such approaches ignore the inherent dependencies and biological structure within these related data.

A joint mixture model is proposed that integrates information from the two data types at the modelling stage to capture their inherent dependency structure, enabling 
simultaneous identification of DMCs and DEGs. The model leverages a joint likelihood function that accounts for the nested structure in the data, with parameter estimation performed using an expectation-maximisation algorithm.

Performance of the proposed method, \texttt{idiffomix}, is assessed through a thorough simulation study and the approach is used to analyse RNA-Seq and DNA methylation array data from matched healthy and breast invasive carcinoma tumour samples from The Cancer Genome Atlas (TCGA) project.
Several genes, identified as non-differentially expressed when the data types were modelled independently, had high likelihood of being differentially expressed when associated methylation data were integrated into the analysis. 
Subsequent gene ontology analysis indicated that many of the differentially methylated CpG sites and differentially expressed genes identified by the joint mixture model are involved in important, in some cases cancer related, biological processes and pathways. 

The \texttt{idiffomix} approach highlights the advantage of an integrated analysis via a joint mixture model over independent
analyses 
of the two data types; genome-wide and cross-omics information is simultaneously utilised 
providing a more comprehensive view. An open source R package is available 
to facilitate widespread use of \texttt{idiffomix}.

\end{abstract}

\section{Introduction}
The epigenetic process of methylation/demethylation of a cytosine-guanine dinucleotide (CpG) site in DNA is an important biomarker in cancer studies \citep{Berger}. This addition or removal of a methyl group from the C5 of a cytosine ring is a heritable change that does not result in DNA sequence alteration \citep{Moore}. Gene promoter regions are the DNA sequences located in the upstream direction, i.e. $5'$ direction of the transcription start site (TSS). Methylation in these promoter regions can block the binding of necessary proteins and transcription factors to the DNA, preventing initiation of transcription of the gene to messenger RNA (mRNA). Methylated DNA is also known to modify histones to produce compact chromatin structure making the DNA less accessible for transcription \citep{Moore}. As DNA methylation makes DNA less accessible for transcription, it thus plays a significant role in controlling gene expression and cellular growth \citep{Bird, suzuki2008dna}. 
Silencing of tumour suppressor genes due to hypermethylation of promoter genes has been studied as an important biomarker in cancer settings \citep{jones2001role,jones2002fundamental}. Genes responsible for cell growth and division, including some oncogenes have been observed to become hyperactive due to hypomethylation \citep{van2017oncogenic}. The relationship between gene expression regulation and DNA methylation has been explored to identify key genes and assess how methylation impacts their function in various neurodegenerative diseases, such as Alzheimer's disease \citep{Alzheimer}. Differential analysis of gene expression and DNA methylation is therefore important for understanding various biological processes and for early detection and treatment of various diseases. Comprehensive measurement of gene expression and DNA methylation has been made possible on a genome wide scale by high-throughput technologies. 

A number of statistical models have been proposed to model gene expression and DNA methylation datasets separately in order to respectively identify differentially expressed genes (DEGs) and differentially methylated CpG sites (DMCs). For gene expression data, models like \texttt{limma-voom}, \texttt{edgeR} and \texttt{DESeq2} employ data transformation for empirical Bayesian modelling or negative binomial distribution models to identify DEGs \citep{robinson2010edger,law2014voom,love2014moderated}. For DNA methylation analysis, statistical models like \texttt{limma}, \texttt{PanDM}, \texttt{FastDMA} and \texttt{betaclust} have been developed to identify DMCs by employing techniques like empirical Bayesian modelling, nonparametric tests and mixture modelling \citep{FastDMA,limma, PanDM,Majumdar}.
\cite{xu2019integrative} identify DEGs and DMCs independently and subsequently employ a correlation analysis to study the associations between them while \cite{xie2011integrative} identified DEGs and employed $t$-tests and regression analysis to infer if changes in methylation were correlated to expression change. \cite{alivand2021integrative} used network analysis to combine DEGs and DMCs to study their association. To find tumour specific candidate genes, \cite{wang2018integrated} examined the association between differentially methylated regions (DMRs) and DEGs. In these ``two-step'' approaches, subtle yet biologically relevant associations between the two data types can be missed when they are modelled independently; integrating the data through a single model would allow simultaneous identification of DEGs and DMCs, providing a more comprehensive analysis of gene regulation being affected by aberrant methylation. 

An integrated approach to modelling gene expression and DNA methylation data was proposed by  \cite{spainhour2019correlation} where Pearson's correlation was employed to study the associations between the two data types . The Bioconductor package \texttt{iNETgrate}  \citep{iNETgrate}  incorporates the two data types at the modelling stage by combining them to form a single gene network. The \texttt{INTEND} model \citep{itai2023integration} uses Lasso regression to predict the gene expression level based on associated methylation values. \cite{kormaksson2012integrative} proposed an integrated mixture model to cluster omics data independently as well as jointly to identify biologically meaningful clusters. The \texttt{EBADIMEX} method \citep{madsen2019ebadimex} proposes moderated $t$-tests and $F$-tests with empirical Bayes priors to integrate the two data types for differential analysis and sample classification while the \texttt{BioMethyl} package \citep{wang2019biomethyl} uses linear regression models to analyse the impact of each CpG site's methylation on gene expression. \cite{jeong2010empirical} proposed 
an integrated empirical Bayes model using marginal linear models to integrate the two data types and applied several user-defined thresholds to classify the genes. While these  models jointly consider gene expression and methylation array data, they typically operate at the gene level, neglecting the nested dependency structure between CpG sites and their corresponding genes. 
There is therefore a need to simultaneously analyse both data types, while accounting for their inherent dependency structure, when identifying DEGs and DMCs. Specifically, the nested dependency structure, created by mapping CpGs to genes based on their positions within promoter regions, needs to be accounted for -- while some genes may experience differential methylation in associated CpG sites without corresponding differential expression, others could exhibit differential expression due to epigenetic factors other than methylation.

Here, a joint mixture model approach, termed \texttt{idiffomix}, is proposed for the integrated differential analysis of gene expression data and methylation array data that accounts for their nested dependency structure. A key dependency matrix parameter is used in the joint mixture model to allow the methylation state of a CpG site
to depend on the expression level status of the gene to which it is associated. The model parameters are estimated via an EM algorithm \citep{EM}. In the E-step, as the expected values of the latent variables  are intractable, approximate but tractable estimates are employed. As chromosomes can be assumed to be independent, the joint mixture model is parallelized across chromosomes to enhance computational efficiency.
The performance of  \texttt{idiffomix} is assessed through a thorough simulation study and application to a publicly available breast cancer dataset where the data are derived from bulk tissue. Several genes implicated in breast cancer, identified to be non-DEGs when the two data types are modelled independently, are 
inferred to be differentially expressed when their corresponding CpG sites' methylation is taken into consideration. Further, gene enrichment analysis revealed the identified DEGs and DMCs to be associated with significant biological processes and  several cancer related pathways, e.g., \textit{MAPK cascade}, \textit{cAMP signaling pathway}, and \textit{Hippo signaling pathway}. To facilitate widespread implementation of \texttt{idiffomix}, an open source R package is available at \href{https://github.com/koyelucd/idiffomix}{Github}.

\section{Gene expression and methylation data}
\label{sec2}

\subsection{Quantifying gene expression and DNA methylation}
Gene expression levels can be quantified as count values using RNA sequencing (RNA-Seq) technology
\citep{wang2009rna} where the count values represent the number of sequencing reads that align to each gene. An increase in gene expression count between biological conditions suggests upregulation of a gene, while a decrease in count suggests downregulation. Differences in gene expression count levels between biological conditions are analysed to identify DEGs. For accurate comparison of gene expression between conditions, RNA-Seq data needs to be normalised to account for differences in library size, which refers to the total number of reads sequenced for each sample. Library size scaling factors are calculated using the Trimmed Mean of M-values approach \citep{robinson2010edger}, and raw counts for each gene are divided by the normalized library sizes and multiplied by one million to get counts per million (CPM) values. Highly expressed genes can increase the variance in the  data therefore the CPM values are typically transformed to log counts per million (log-CPM) to stabilise the variance and facilitate analysis using Gaussian models \citep{law2014voom}.

Methylation arrays quantify the level of methylation at a CpG site as a \textit{beta} value \citep{Du}. These values are continuous and bounded in nature ranging from $0$ to $1$, where a value close to $0$ indicates hypomethylation, a value close to $0.5$ indicates hemimethylation and a value close to $1$ indicates hypermethylation at a CpG site. Differences in methylation levels are analysed across different biological conditions to identify DMCs. For analysis, as bounded \textit{beta} values violate Gaussian assumptions,  typically \textit{beta} values are transformed using a \textit{logit} transformation to produce \textit{M}-values. Such a transformation results in real valued data 
which is more appropriate for the application of Gaussian models.

From each RNA sample, gene expression levels of $G$ genes are quantified while from each DNA sample, methylation levels at $C$ CpG sites are quantified, where typically $C \gg G$.  The RNA-Seq and methylation array data, each collected from the same set of individuals, exhibit a many-to-one mapping, in that each CpG site can be assigned to a single gene, although some genes may have no associated CpG sites. This nested structure results in a complex relationship between gene expression and methylation patterns, as variations in methylation levels at multiple CpG sites within a genomic locus may collectively be associated with variations in gene expression.

\subsection{Motivating breast cancer study}

\label{idiffomixsec:data}

Gene expression and methylation data in matched healthy and breast invasive carcinoma tumour samples from The Cancer Genome Atlas (\href{https://portal.gdc.cancer.gov/projects/TCGA-BRCA}{TCGA}) are are publicly available in the Genomic Data Commons (\href{https://portal.gdc.cancer.gov/}{GDC}) data portal. 
Here, a sub-selection of data from $N = 5$ patients with the \textit{duct and lobular neoplasm} disease type and for whom both gene expression and methylation array data were available, were analysed. The donor ID (case ID in parenthesis) of patients selected are: DO1253 (TCGA-E9-A1NG), DO1254 (TCGA-BH-A0AU), DO1257 (TCGA-BH-A0DG), DO1283 (TCGA-BH-A1EU) and DO1299 (TCGA-BH-A0BM).
For each patient, RNA-Seq data along with 450K methylation array data were obtained
from normal (herein benign) and primary tumour breast tissues using Illumina HiSeq Sequencing and Illumina HumanMethylation450 BeadChip array \citep{steemers2005illumina}, respectively.
The data are derived from bulk tissue rather than single-cell samples, capturing the aggregate molecular signals across all cell types within the tissue. The RNA-Seq data comprised of gene expression levels for $ 20,499$ genes while the methylation data comprised of methylation levels for $394,356$ CpG sites.

The RNA-Seq data consisted of raw counts depicting the gene expression levels. To ensure data quality, only genes whose sum of expression counts across both biological conditions $> 5$ were retained, resulting in $G = 15,722$ genes. The data were normalized to account for 
differences in library sizes. The normalized count data were used to obtain CPM values which were further log-transformed to obtain log-CPM values. Given the paired design of the motivating setting, the log-fold changes between the tumour and benign samples were calculated for each gene in every patient and used in the subsequent analyses. For the methylation array data, 
CpG sites located within promoter regions were selected, specifically those within 200 nucleotides of the transcription statrt site (TSS200), the first exon, or the $5'$ untranslated region (UTR) region, resulting in $C = 94,873$ CpG sites. The \textit{beta} values at these CpG sites were then \textit{logit} transformed to \textit{M}-values. Similar to the RNA-Seq data, given the paired design, the difference in \textit{M}-values
between tumour and benign samples were calculated for each CpG site in every patient and used in the subsequent analyses.

\section{The idiffomix method}

\subsection{A joint mixture model for gene expression and methylation
data}
\label{sec:model}

Given the paired experimental design of the motivating setting, 
 the log-fold change of the normalized RNA-Seq data between the two biological conditions for the $g^{th}$ gene is denoted $\bm{x}_{g}$. Specifically, $\bm{x}_g =  ({x}_{g1}, \ldots, x_{gn}, \ldots, {x}_{gN})$, where ${x}_{gn}$ signifies the log-fold change in gene expression levels for the $g^{th}$ gene  from the $n^{th}$ patient's RNA sample. The collection of log-fold change values for all $G$ genes and $N$ patients is denoted $\bm{X} = (\bm{x}_{1}, \ldots, \bm{x}_{g}, \ldots, \bm{x}_{G})$, a 
$G \times N$ matrix.
Similarly, the difference in \textit{M}-values for CpG site $c$ located on gene $g$ between the two biological conditions is denoted 
${\bm{y}}_{gc} = ({y}_{gc1}, \ldots, y_{gcn}, \ldots, {y}_{gcN})$, where ${y}_{gcn}$ represents the 
difference in \textit{M}-values at CpG site $c$, located on gene $g$ from patient $n$'s DNA sample.  Each gene $g$ has $C_g$ associated CpG sites such that $C = \sum_{g=1}^G C_g$. It is possible for a gene to have no associated CpG sites, in which case $C_g = 0$ for that gene. The collection of \textit{M}-value differences between the two biological conditions for all $C$ CpG sites  and $N$ patients is denoted $\bm{Y} = (\bm{y}_{11}, \ldots, \bm{y}_{1C_1}, \ldots, \bm{y}_{g1}, \ldots, \bm{y}_{gC_g}, \ldots, \bm{y}_{G1}, \ldots, \bm{y}_{GC_G 
})$, a $C \times N$ matrix.

To simultaneously identify DEGs and DMCs, while accounting for the nested structure between genes and CpG sites, a joint mixture modelling approach, termed \texttt{idiffomix}, is proposed. 
Expression levels at each gene are assumed
to undergo one of three possible state changes between the benign and tumour conditions: if expression levels decrease between the tumour and matched normal samples, the gene is considered downregulated (E-) and will have a large negative log-fold change. Conversely, if expression levels increase 
in the tumour sample, the gene is deemed upregulated (E+) with large positive log-fold change value. If no change is observed, a gene is categorized as non-differentially expressed (E0), with log-fold change values close to 0. Thus, under \texttt{idiffomix}, the log-fold changes in the RNA-Seq data are assumed to have been generated from a heterogeneous population modelled by a $K = 3$ component mixture model.
Similarly, methylation levels at a CpG site are also assumed to undergo one of three possible state changes between the two biological conditions: a CpG site is considered hypomethylated (M-) if the methylation level decreases between the tumour and benign samples, resulting in large negative differences in \textit{M}-values. A CpG site is deemed to be hypermethylated (M+) if methylation increases in the tumour sample compared to the benign sample, represented by large positive differences in \textit{M}-values. A CpG site will be non-differentially methylated (M0) if the difference in \textit{M}-values between the two conditions is close to 0. Thus, the differences in \textit{M}-values are assumed to be generated from an $L = 3$ component mixture model.

As is typical in mixture models, an incomplete data approach is employed to facilitate inference. 
For gene $g$, an indicator variable $u_{gk}$ is introduced, where $u_{gk} = 1$ if the gene $g$ belongs to cluster $k$, for $g = 1, 2, \ldots, G$ and $k = 1, 2, ..., K$. The collection of indicator variables for $G$ genes is denoted $\bm{U} = (\bm{u}_1, \ldots, \bm{u}_g, \ldots, \bm{u}_G)$, a $G \times K$ matrix. Further, within each component, the log-fold change data are assumed to be independent and identically Gaussian distributed, i.e., ${x}_{gn} | (u_{gk} = 1) \sim N(\mu_k, \sigma_k^2)$, for $g = 1, 2, \ldots, G$, $n = 1,2,\ldots, N$ and $k = 1, 2, \ldots, K$. Similarly, for CpG site $C$, an indicator variable $v_{gcl}$ is introduced, where $v_{gcl} = 1$ if CpG site $c$, located on gene $g$, belongs to cluster $l$. The collection of indicator variables for $C$ CpG sites is denoted $\bm{V} = (\bm{v}_{11}, \ldots, \bm{v}_{1C_1}, \ldots, \bm{v}_{g1}, \ldots, \bm{v}_{gC_g}, \ldots, \bm{v}_{G1}, \ldots, \bm{v}_{GC_G})$, a $C \times L$ matrix.
The differences in \textit{M}-values are also assumed to be independent and identically Gaussian distributed within a component, i.e., $ {{y}}_{gcn}|(v_{gcl}=1) \sim N(\lambda_l,\rho_{l}^2)$ for $g = 1, 2, \ldots, G$, $c = 1, 2, \ldots, C_g$, $n = 1,2,\ldots,N$ and $l = 1, 2, \ldots, L$.

\begin{figure}[ht]
    \centering
\begin{tikzpicture}[
    node distance=0.5cm and 2cm,
    every node/.style={font=\small},
    box/.style={draw, rectangle, minimum width=1cm, minimum height=0.7cm, align=center},
    circle_node/.style={draw, circle, minimum width=1cm, align=center},
    arrow/.style={-Stealth, thick}
]
\node[box] (y11) {$\bm{y}_{11}$};
\node[box, below=of y11] (y12) {$\bm{y}_{12}$};
\node[box, below=of y12] (y13) {$\bm{y}_{13}$};
\node[box, below=of y13] (y21) {$\bm{y}_{21}$};
\node[box, below=of y21] (y22) {$\bm{y}_{22}$};
\node[below=of y22,below=1.8cm of y22](y01) {};
\node[draw, fit={(y11) (y01)}, inner sep=0.5cm, minimum width=2.5cm, minimum height=8cm, label=above:{\parbox{2.2cm}{\begin{center}Methylation \\ data \end{center}}}, yshift=.26cm] (methyl_block) {};

\node[circle_node, right=of y11] (v11) {$\bm{v}_{11}$};
\node[circle_node, right=of y12] (v12) {$\bm{v}_{12}$};
\node[circle_node, right=of y13] (v13) {$\bm{v}_{13}$};
\node[circle_node, right=of y21] (v21) {$\bm{v}_{21}$};
\node[circle_node, right=of y22] (v22) {$\bm{v}_{22}$};
\node[below=1.5cm of v22](v01) {};
\node[draw, fit={(v11) (v01)}, inner sep=0.5cm, minimum width=2.5cm, minimum height=8cm, label=above:{\parbox{2.2cm}{\begin{center}CpG site \\ latent variables\end{center}}}, yshift=.11cm] (cpg_block) {};

\node[circle_node, right=of v12] (u1) {$\bm{u}_{1}$};
\node[circle_node, below=1.6cm of u1] (u2) {$\bm{u}_{2}$};
\node[circle_node, below=1cm of u2] (u3) {$\bm{u}_{3}$};

\node[draw, fit={(u1) (u3)}, inner sep=0.5cm, minimum width=2.5cm, minimum height=8.6cm, label=above:{\parbox{2.2cm}{\begin{center}Gene \\latent variables\end{center}}}, yshift=.3cm] (genes_block) {};

\node[box, right=of u1] (x1) {$\bm{x}_{1}$};
\node[box, right=of u2] (x2) {$\bm{x}_{2}$};
\node[box, right=of u3] (x3) {$\bm{x}_{3}$};

\node[draw, fit={(x1) (x3)}, inner sep=0.5cm, minimum width=2.5cm, minimum height=8.6cm, label=above:{\parbox{2.2cm}{\begin{center}Gene \\expression data\end{center}}}, yshift=.3cm] (gene_expr_block) {};

\draw[->] (v11) -- (y11);
\draw[->] (v12) -- (y12);
\draw[->] (v13) -- (y13);
\draw[->] (v21) -- (y21);
\draw[->] (v22) -- (y22);

\draw[->] (u1) -- (v11);
\draw[->] (u1) -- (v12);
\draw[->] (u1) -- (v13);
\draw[->] (u2) -- (v21);
\draw[->] (u2) -- (v22);

\draw[->] (u1) -- (x1);
\draw[->] (u2) -- (x2);
\draw[->] (u3) -- (x3);

\end{tikzpicture}
    \caption{\textbf{Graphical model of \texttt{idiffomix}.}\\
\small The joint mixture model for integrated differential analysis of gene expression and methylation data. Here, for example, gene expression data arises from 3 genes: gene 1 has 3 associated CpG sites while gene 2 has 2 associated CpG sites. Gene 3 has no associated CpG sites.}

   \label{idiffomix_fig:graphical_model}
\end{figure}

 To account for the nested dependency structure between genes and CpG sites, the two mixture models are integrated. Firstly, the proportion of genes belonging to  each cluster is denoted as $\bm{\tau} = (\tau_1, \ldots, \tau_k, \ldots, \tau_K)$. The dependencies between the genes and CpG sites are then accounted for through a key $L \times K$ matrix parameter $\bm{\pi}$. The value $\pi_{l|k}$ is the probability of a CpG site belonging to cluster $l$, given its associated associated gene belongs to cluster $k$, 
 for $l = 1, 2, \ldots, L$ and $k = 1, 2, \ldots, K$. 
 Figure \ref{idiffomix_fig:graphical_model} provides a graphical model of the \texttt{idiffomix} model structure.

 Denoting the parameters in the gene expression mixture model collectively as $\bm{\theta} = (\bm{\theta}_1, \ldots, \bm{\theta}_K)$, where $\bm{\theta}_{k} = $ (${\mu}_{k}$, ${\sigma}_{k}^2$), and the parameters in the methylation data clusters collectively as $\bm{\phi} = (\bm{\phi}_1, \ldots, \bm{\phi}_L)$, where $\bm{\phi}_{l} = $ (${\lambda}_{l}$, ${\rho}_{l}^2$), the joint probability density function for the log-fold change RNA-Seq data $\bm{X}$, differences in \textit{M}-value methylation data $\bm{Y}$ and latent variables $\bm{U}$ and $\bm{V}$ is,
\begin{equation}
\begin{aligned}
P({\bm{X}},{\bm{Y}},\bm{U},\bm{V}|\bm{\tau},\bm{\pi}, \bm{\theta}, \bm{\phi}) = & 
    \prod\limits_{g=1}^G \left\{ \prod\limits_{k=1}^K P({\bm{x}}_g|\bm{\theta}_k)^{u_{gk}} \prod\limits_{c=1}^{C_g} \prod\limits_{l=1}^L P({\bm{y}}_{gc}|\bm{\phi}_{l})^{v_{gcl}} \right\} \\ & \times \prod\limits_{g=1}^G  \prod\limits_{k=1}^K \left\{\tau_k \prod\limits_{c=1}^{C_g} \prod\limits_{l=1}^{L} \pi_{l|k}^{v_{gcl}} \right\}^{u_{gk}}.
    \end{aligned}
    \label{idiffomix_eqn:joint_model}
\end{equation}
 If a gene has no associated CpG sites then the products over $c$ and $l$ in (\ref{idiffomix_eqn:joint_model}) are equal to one. From (\ref{idiffomix_eqn:joint_model}), it is clear that if $\pi_{l|k} = \pi_{l|k'}$ for all $k, k'$, then the status of CpG sites and genes are independent. In such a case, the model is equivalent to two independent mixture models.

Assuming that the data from the $N$ patients are conditionally independent given their component membership then the probability of the RNA-Seq data is 
$P(\bm{x}_g | \bm{\theta}_k) = \prod_{n=1}^{N} p(x_{gn} |\bm{\theta}_k)$,
and similarly, for the methylation data, 
$P(\bm{y}_{gc} | \bm{\phi}_{l}) = \prod_{n=1}^{N} p(y_{gcn} |\bm{\phi}_{l})$.
Thus the complete data log-likelihood function 
for the joint mixture model is,
\begin{equation}
\begin{aligned}
    \ell_C(\bm{\tau},\bm{\pi}, \bm{\theta}, \bm{\phi}| {\bm{X}}, {\bm{Y}}, \bm{U}, \bm{V}) = & \sum\limits_{g=1}^G \sum\limits_{k=1}^K  \sum\limits_{n=1}^N u_{gk}\log p({x}_{gn}|\bm{\theta}_k)   \\ & + \sum\limits_{g=1}^G\sum\limits_{c=1}^{C_g} \sum\limits_{l=1}^L \sum\limits_{n=1}^N  \; v_{gcl}\log p({y}_{gcn}|\bm{\phi}_{l})    \\
    & + \sum\limits_{g=1}^G \sum\limits_{k=1}^K u_{gk}\log \tau_k + \sum_{g=1}^{G}\sum_{k=1}^{K}\sum\limits_{c=1}^{C_g} \sum\limits_{l=1}^L u_{gk}v_{gcl}\log\pi_{l|k}.
    \end{aligned}
    \label{idiffomix_eqn:complete_llk}
\end{equation}         
\subsection{Inference via an EM algorithm}

The expectation-maximisation (EM) algorithm \citep{EM} is used to compute, from  (\ref{idiffomix_eqn:complete_llk}), the maximum likelihood estimates of the 
 model parameters, i.e., $\hat{\bm{\tau}}$, $\hat{\bm{\pi}}$, $\hat{\bm{\theta}}$, and $\hat{\bm{\phi}}$ and the expected values of the latent variables, conditional on the current parameter estimates and the observed data.

At the E-step of the EM algorithm, the expected value of the complete data log-likelihood function (\ref{idiffomix_eqn:complete_llk}) with respect to the conditional distributions of the latent variables is calculated, given the observed data and the current estimates of the model parameters. The required expected values are therefore,
\begin{eqnarray*}\nonumber
\hat{u}_{gk} & = & \mathbb{E}(u_{gk}| {\bm{x}}_g, \bm{\tau}_k, \bm{\pi}_{l|k}, \bm{\theta}_k),\\\label{eqn:latentvars}
\hat{v}_{gcl} & = & \mathbb{E}(v_{gcl}| {\bm{y}}_{gc}, \bm{\pi}_{l|k}, \bm{\phi}_l) \\\nonumber \widehat{u_{gk}v_{gcl}}  & = & \mathbb{E}(u_{gk}v_{gcl}| {\bm{x}}_g, {\bm{y}}_{gc}, \bm{\tau}_k, \bm{\pi}_{l|k}, \bm{\theta}_k, \bm{\phi}_l ),
\end{eqnarray*}
for $g = 1, 2, \ldots, G, k = 1, 2, \ldots, K, c = 1, 2, \ldots, C_g$ and $l = 1, 2, \ldots, L$.
While these are intractable, a tractable approximation is available by considering the conditional expected values of each latent variable given the other and employing an algorithm similar to that proposed in \cite{BMMSTcoordinateascent} and \cite{coordinateascent} at the E-step (see Appendix~A.1 for full details). For the M-step, the expected complete data log-likelihood function is maximised with respect to the model parameters $\bm{\tau}$, $\bm{\pi}$, $\bm{\theta}$ and $\bm{\phi}$;
closed form solutions are available and full details  are given in Appendix~A.1.

To initialise the EM algorithm, here the latent variables $\bm{U}$ and $\bm{V}$ are initialised. Clustering algorithms such as $k$-means or \texttt{mclust} \citep{mclust5} could be used to independently cluster the two data types into $K=3$ and $L=3$ clusters. Here we instead chose to employ a quantile initialization approach to ensure that the initial clusters were well separated. 
The quantile initialization approach was applied to the two data types independently such that the E- and M- clusters consisted of genes and CpG sites, respectively, with the lowest $10\%$  differential expression and methylation levels, the E+ and M+ clusters contained genes and CpG sites, respectively, with the highest $10$\% of differential expression and methylation levels and the remaining genes and CpG sites were allocated to clusters E0 and M0 respectively. Given the resulting initialised values of $\bm{U}$ and $\bm{V}$, an M-step was used to calculate  starting values for the model parameters ${\bm{\tau}}$, ${\bm{\pi}}$, $\bm{\theta}$, and $\bm{\phi}$.  The E-step and M-step were then iterated until convergence which here was deemed to be achieved when the absolute change in all parameter estimates between successive iterations was less than the threshold of $1 \times 10^{-5}$, to yield estimates $\bm{\hat{\tau}}$, $\bm{\hat{\pi}}$, $\bm{\hat{\theta}}$ and $\bm{\hat{\phi}}$.

On convergence of the EM-algorithm, the estimates of the latent variables  $\bm{U}$ and $\bm{V}$ provide the posterior probabilities of cluster membership for genes and CpG sites respectively. 
Cluster assignment is then performed using the maximum a posteriori (MAP) rule, where each gene or CpG site is assigned to the cluster for which it has highest posterior probability of membership. Thus, DEGs are deemed to be those assigned to clusters E- and E+ while DMCs are deemed to be those assigned to clusters M- and M+,  based on the joint analysis of both datasets. The uncertainty of gene $g$'s cluster assignment is available as $1 - \max\limits_{k = 1, \ldots, K}(\hat{u}_{gk})$ while the uncertainty of CpG site $c$'s cluster assignment is available as $1 - \max\limits_{l = 1, \ldots, L}(\hat{v}_{gcl})$, providing deeper insight than available via a hard clustering approach.

Due to independence of chromosomes and to ease the computational burden, the model is fitted to each chromosome independently in parallel.
Assuming a genome has $J$ chromosomes the joint mixture model in (\ref{idiffomix_eqn:joint_model}) is fitted separately $J$ times, once for each chromosome $j = 1, 2, \ldots, J$. In each instance the model is fitted to the $G_j$ genes and $C_j$ CpG sites
associated with gene $j$. 

\subsection{A generalised joint mixture model}
While the joint mixture model in (\ref{idiffomix_eqn:joint_model}) was formulated based on the paired experimental design of the motivating study outlined in Section \ref{idiffomixsec:data}, the \texttt{idiffomix} method can be generalised to other experimental designs and omic datasets, for example where two omics data types are collected from the same individuals in $R \ge 2$ different biological conditions or time points. 
In such a generalized scenario, while the observed omics data may be transformed to e.g., achieve Gaussianity or stabilise variances, the (transformed) observed data, rather than fold changes or differences between the $R$ replicates, could be modelled directly.
In this generalised \texttt{idiffomix} model, the RNA-Seq data, say, at gene $g$ are denoted $\bm{x}_g = (x_{g11}, \ldots, x_{gnr}, \ldots, x_{gNR})$, where $x_{gnr}$ represents the (transformed) gene expression level at the $g^{th}$ gene, collected from the $n^{th}$ patient's $r^{th}$ biological condition or time point. Similarly, at the $c^{th}$ CpG site the methylation data $\bm{y}_{gc} = (y_{gc11}, \ldots, y_{gcnr}, \ldots, y_{gcNR})$ where $y_{gcnr}$ denotes the (transformed) \textit{beta} value at the $c^{th}$ CpG site located on the $g^{th}$ gene, collected from the $n^{th}$ patient's $r^{th}$ biological condition or time point. The values of $K$ and $L$ may vary depending on the study objective, similar to the approach taken in \cite{Majumdar}. Finally, while here the \texttt{idiffomix} model assumes Gaussian distributions for $P(\bm{x}_g)$ and $P(\bm{y}_{gc})$, this need not be the case and other distributions could be specified as appropriate for the data and research question at hand, with or without data transformations as needed.

\section{Results}
\subsection{Simulation study}

\subsubsection{Data generation}
To assess the performance of \texttt{idiffomix}, a simulation study was conducted where the simulation settings were chosen to mirror those in the motivating breast cancer study data. One hundred RNA-Seq and methylation array datasets were simulated based on (\ref{idiffomix_eqn:joint_model}) from two biological conditions (conditions A and B), from $N = 4$ patients and $K = L = 3$. For the simulated RNA-Seq data, $G = 500$ genes. Based on the TCGA data, the number of CpG sites linked to each gene ranged from $1$ to $100$, with $95\%$ of genes having fewer than $30$ CpG sites. Therefore, for the simulated methylation data, the number of CpGs linked to each gene was drawn from a uniform distribution $U(3, 30)$
resulting in the simulated methylation datasets having $5,000$ CpG sites, on average.

The RNA-Seq raw counts under condition A were generated from a negative binomial distribution with mean of $10,000$ and dispersion parameter of $5$.
Ten percent of the $G = 500$ genes were simulated to be downregulated (i.e., in cluster E-) and raw counts for these genes in condition B were simulated from a negative binomial distribution with mean and dispersion parameters of $4,000$ and $5$ respectively.
Similarly, $10\%$ of the genes were simulated to be upregulated (i.e., in cluster E+) 
with counts generated for condition B from a negative binomial with mean  and dispersion parameters of $60,000$ and $5$ respectively. For non-differential genes (i.e., those in cluster E0), raw counts for condition B were generated as for condition A, i.e., from a negative binomial distribution with mean and dispersion parameters of $10,000$ and $5$ respectively. To emulate the characteristics of the real data, random Poisson noise was added to each generated count, where the Poisson mean  was the generated count value.

Given the nested structure of CpG sites on genes, simulation of the methylation data depends on the cluster membership of CpG sites' associated genes, as quantified by $\bm{\pi}$. To  assess the robustness of the model across different dependency scenarios, methylation data were generated under three different settings of $\bm{\pi}$, as detailed in Table \ref{idiffomix_tab:combined_probabilities}.
In case 1, probabilities are similar to those resulting from fitting the joint mixture model to the breast cancer dataset. Case 2 considers a high level of dependency,
 represented by 
 a diagonal-heavy $\bm{\pi}$ matrix. In case 3, no dependence between the two data types is assumed.
\begin{table}[htb]
        \caption{Three settings of $\bm{\pi}$ considered in the simulation study.}
    \label{idiffomix_tab:combined_probabilities}
    \begin{tabular}{c}
    \begin{subtable}{0.32\textwidth}
        \centering
        \captionsetup{labelformat=empty}
         \caption{(a) Case 1}
        \begin{tabular}{c|ccc}
            & E- & E0 & E+ \\
            \hline
            M+  & 0.4  & 0.05 & 0.1 \\
            M0 & 0.5  & 0.9  & 0.5 \\
            M-  & 0.1  & 0.05 & 0.4
        \end{tabular}
    \end{subtable}
    \hfill
    \begin{subtable}{0.32\textwidth}
        \centering
        \captionsetup{labelformat=empty}
                \caption{(b) Case 2}
        \begin{tabular}{c|ccc}
           & E- & E0 & E+ \\
            \hline
            M+  & 0.8  & 0.1  & 0.1 \\
            M0 & 0.1  & 0.8  & 0.1 \\
            M-  & 0.1  & 0.1  & 0.8
        \end{tabular}
    \end{subtable}
    \hfill
    \begin{subtable}{0.32\textwidth}
        \centering
        \captionsetup{labelformat=empty}
                \caption{(c) Case 3}
        \begin{tabular}{c|ccc}
            & E- & E0 & E+ \\
            \hline
            M+  & 0.2  & 0.6  & 0.2 \\
            M0 & 0.2  & 0.6  & 0.2 \\
            M-  & 0.2  & 0.6  & 0.2
        \end{tabular}
    \end{subtable}
        \end{tabular}
        \captionsetup{justification=centering, font=small}
        \small Values represent the probabilities of a CpG site belonging to cluster M+, M0 or M-, 
        conditional of their associated gene belonging to cluster E-, E0 or E+.
\end{table}

Given the cluster membership of a CpG site's associated gene, if a CpG site is assumed to be hypermethylated (i.e., belongs to cluster M+) in condition B compared to A, \textit{beta} values were generated from a Beta$(3,20)$ distribution for condition A and from a Beta$(20,3)$ for condition B. Similarly, if the CpG site is hypomethylated (i.e. belongs to cluster M-) in condition B compared to A, \textit{beta} values were generated from Beta$(20,3)$ for condition A and Beta$(3,20)$ for condition B. Non-differentially methylated CpG sites arise when they are either hypermethylated in both conditions, hypomethylated in both conditions or hemimethylated in both conditions. Therefore, if a CpG site is assumed to be non-differential (i.e., belongs to cluster M0), the \textit{beta} values for both conditions A and B were generated from either Beta$(3,20)$ (both hypomethylated), Beta$(20,3)$ (both hypermethylated) or Beta$(4,3)$ (both hemimethylated) distributions.
Zero centred Gaussian noise with standard deviation 0.05 was added to each of the generated \textit{beta} values to emulate the variation in the breast cancer data. 

For analysis, the simulated RNA-Seq data were normalized, log-transformed to log-CPM values and log-fold changes from condition A to B calculated. Similarly, differences in logit-transformed \textit{beta} values  between conditions A and B were calculated for all CpG sites. The \texttt{idiffomix} joint mixture model was then fitted to these simulated log-fold change RNA-Seq values  and to the differences in the simulated methylation \textit{M}-values.

\subsubsection{Simulation study results}

Results under the \texttt{idiffomix} approach were compared to those obtained when the two simulated data types were analysed independently using the \texttt{mclust} and \texttt{limma} approaches \citep{limma,mclust5} . For  \texttt{mclust},  three clusters and the \texttt{EEI}, \texttt{EVI} and \texttt{EII} models were considered and cluster assignment was via the MAP rule.
For \texttt{limma}, a threshold of $0.05$ for Benjamini-Hochberg adjusted $p$-values was used to identify significant differential expression and methylation. 
Performance of the three methods was evaluated based on false discovery rate (FDR), sensitivity, specificity and adjusted Rand index (ARI) \citep{ARI}.

Table \ref{idiffomix_tab:case1_results} details the performance metrics across the 100 simulated datasets generated given $\bm{\pi}$ from case 1 in Table \ref{idiffomix_tab:combined_probabilities}. In terms of identifying DEGs, \texttt{idiffomix} outperforms \texttt{mclust} and \texttt{limma} with lower mean FDR and higher mean sensitivity, specificity and ARI values. 
Similarly when identifying DMCs, \texttt{idiffomix} performs well overall, with only \texttt{limma} having a slightly higher mean sensitivity. Interestingly, \texttt{idiffomix} has much stronger performance than \texttt{mclust} and \texttt{limma}  when identifying DEGs, while all 3 methods show similar performance when identifying DMCs; when information from CpG sites associated with the genes is appropriately accounted for through the joint mixture model's dependency structure, DEG identification improves.  
Table \ref{idiffomix_tab:case2_results} shows similar performance patterns for data generated using $\bm{\pi}$ under case 2 from Table \ref{idiffomix_tab:combined_probabilities} where the level of dependency is high. Results under case 3 show similar clustering solutions between the three methods suggesting the joint mixture model performs as two independent mixture models; results are provided in
Appendix A.2.

\begin{table}[htb]
    \centering
        \caption{Mean performance metrics for 100 simulated datasets  given $\bm{\pi}$ under case 1 from Table \ref{idiffomix_tab:combined_probabilities}.}\begin{tabular}{c}
        \begin{subtable}[t]{\textwidth}
            \centering
            \captionsetup{labelformat=empty}
                        \caption{(a) DEG identification performance}
            \begin{tabular}{l c c c c }
                \hline
                & FDR & Sensitivity & Specificity & ARI \\
                \hline
                \textbf{\texttt{idiffomix}} & \textbf{0.014} (0.011) & \textbf{0.976} (0.015) & \textbf{0.997} (0.003) & \textbf{0.966} (0.017) \\ 
               \texttt{mclust} & 0.102 (0.049) & 0.873 (0.046) & 0.975 (0.015) & 0.800 (0.041) \\ 
                \texttt{limma} & 0.038 (0.021) & 0.764 (0.064) & 0.993 (0.005) & 0.760 (0.059) \\ \hline
            \end{tabular}
        \end{subtable} \\
        \vspace{0.3cm} 
        \begin{subtable}[t]{\textwidth}
            \centering
            \captionsetup{labelformat=empty}
              \caption{(b) DMC identification performance}
            \begin{tabular}{l c c c c }
                \hline
                & FDR & Sensitivity & Specificity & ARI \\
                \hline
                \textbf{\texttt{idiffomix}} & \textbf{0.016} (0.005) & 0.999 (0.001) & \textbf{0.997} (0.001) & \textbf{0.986} (0.004) \\ 
               \texttt{mclust} & 0.019 (0.006) & 0.999 (0.001) & 0.996 (0.001) & 0.983 (0.005) \\ 
                \texttt{limma} & 0.058 (0.006) & \textbf{1.000} ($<$0.001) & 0.987 (0.002) & 0.948 (0.006) \\ \hline
            \end{tabular}
        \end{subtable}
    \end{tabular}\label{idiffomix_tab:case1_results}
    \small  *Standard deviations in parentheses and the top performing method for each metric is highlighted in boldface.
\end{table}

\begin{table}[htb]
    \centering
        \caption{Mean performance metrics for 100 simulated datasets given $\bm{\pi}$ under case 2 from Table \ref{idiffomix_tab:combined_probabilities}.}\begin{tabular}{c}
        \begin{subtable}[t]{\textwidth}
            \centering
            \captionsetup{labelformat=empty}
                        \caption{(a) DEG identification performance}
            \begin{tabular}{l c c c c }
                \hline
                & FDR & Sensitivity & Specificity & ARI \\
                \hline
                \textbf{\texttt{idiffomix}} & \textbf{0.003} (0.006) & \textbf{0.997} (0.005) & \textbf{0.999} (0.001) & \textbf{0.995} (0.007) \\ 
               \texttt{mclust} & 0.102 (0.049) & 0.873 (0.046) & 0.975 (0.015) & 0.800 (0.041) \\ 
                \texttt{limma} & 0.038 (0.021) & 0.764 (0.064) & 0.993 (0.005) & 0.760 (0.059) \\ \hline
            \end{tabular}
        \end{subtable} \\
        \vspace{0.3cm} 
        \begin{subtable}[t]{\textwidth}
            \centering
            \captionsetup{labelformat=empty}
              \caption{(b) DMC identification performance}
            \begin{tabular}{l c c c c }
                \hline
                & FDR & Sensitivity & Specificity & ARI \\
                \hline
                \textbf{\texttt{idiffomix}} & \textbf{0.009} (0.003) & 1.000 ($<$ 0.001) & \textbf{0.995} (0.002) & \textbf{0.987} (0.004) \\ 
               \texttt{mclust} & 0.011 (0.004) & 1.000 ($<$ 0.001) & 0.994 (0.002) & 0.984 (0.005) \\ 
                \texttt{limma} & 0.050 (0.005) & 1.000 ($<$ 0.001) & 0.973 (0.003) & 0.930 (0.007) \\ \hline
            \end{tabular}
        \end{subtable}
    \end{tabular}\label{idiffomix_tab:case2_results}    
    \small  *Standard deviations in parentheses and the top performing method for each metric is highlighted in boldface.
\end{table}

All computations were performed using R (version 4.3.3) \citep{R2024} on a Windows 11 operating system with an Intel Core i7 CPU (2.70GHz) and 16GB of RAM. In terms of computational cost, for example, fitting \texttt{idiffomix} to one simulated dataset of RNA-Seq and methylation data took 5.34 minutes. To explore the impact of a larger value of $N$ on computational cost, further simulation studies were performed where $N$ varied from $N = 4$ to $N = 200$. The computational cost increased linearly with $N$, consistent with the form of the model’s likelihood function, suggesting a time complexity of $O(N)$. Further details on computational cost are provided in Appendix A.3.

\subsection{Application to breast cancer gene expression and methylation data}

The \texttt{idiffomix} model is fit to the publicly available breast cancer RNA-Seq and methylation array data outlined in Section \ref{idiffomixsec:data}. 
The joint mixture model was applied to data from each chromosome individually; for clarity, results for a single chromosome (chromosome 7) are provided here with results for all other chromosomes provided in Appendix A.4. Analysing the RNA-Seq and methylation data independently indicated large numbers of DMCs and DEGs on chromosome 7 and hence, the results for chromosome 7 are given here.  

 \begin{figure}[H]
    \centering
    \includegraphics[width=\textwidth]{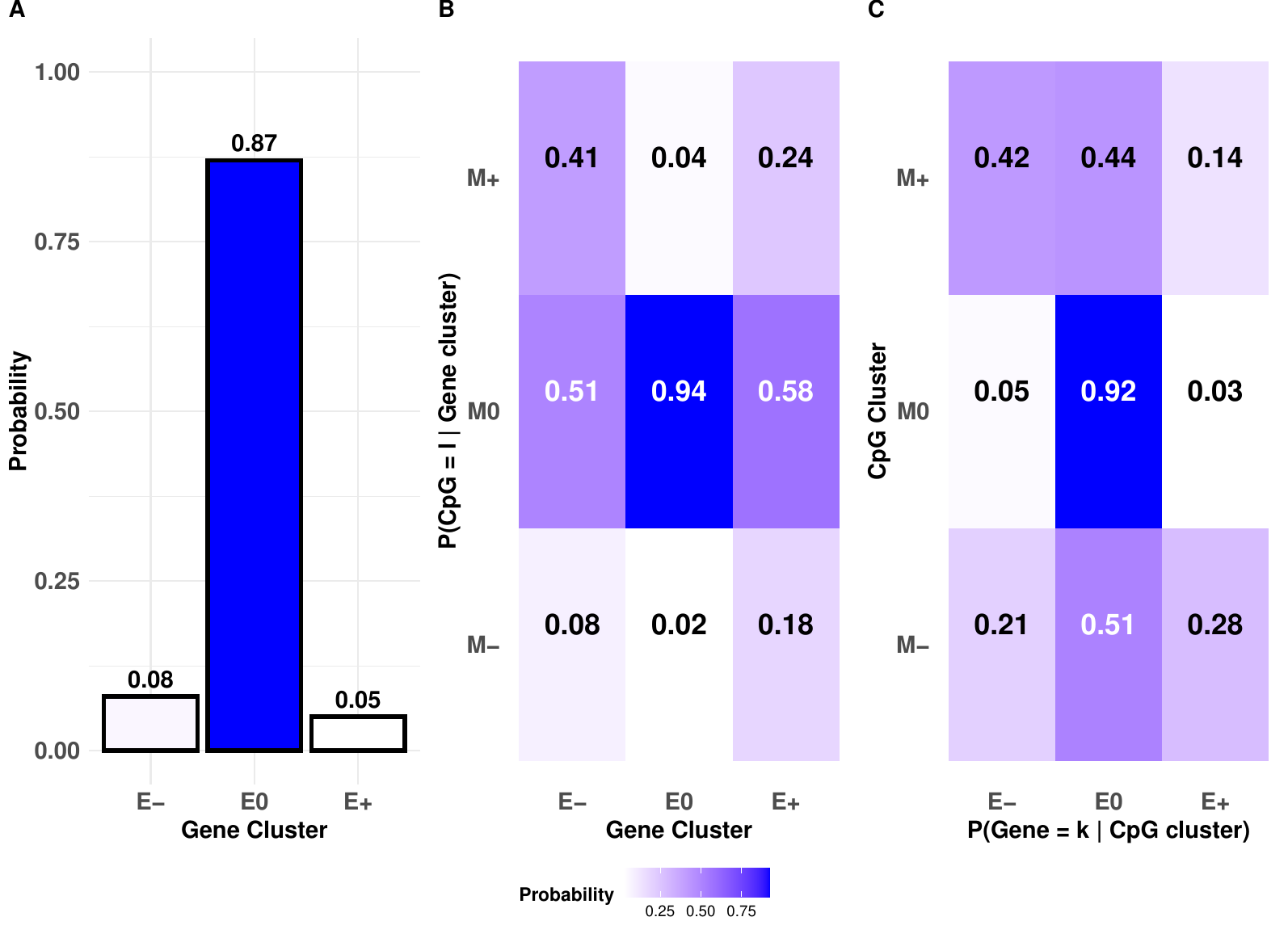}
    \caption{\textbf{ \texttt{Idiffomix} applied to chromosome 7 of TCGA breast cancer data.} \\
    \small (A) estimated cluster  membership probabilities $\hat{\bm{\tau}}$, (B) the estimated matrix $\hat{\bm{\pi}}$ of conditional probabilities of CpG site methylation status given gene cluster membership, (c) conditional probabilities of genes belonging to cluster $k$ given a single CpG site associated with the gene belongs to cluster $l$.}
    \label{idiffomix_fig:real_data_result}
\end{figure}

Figure \ref{idiffomix_fig:real_data_result} panel A displays the probability of a gene in chromosome 7 belonging to each of the E-, E0 or E+ clusters, with a large majority deemed to be non-differentially expressed and allocated to the E0 cluster. Panel B in Figure \ref{idiffomix_fig:real_data_result} details the estimated matrix $\hat{\bm{\pi}}$ of conditional probabilities of a CpG site's cluster membership, given its gene's cluster. 
Interestingly, a CpG site has highest probability of being in cluster M0 (i.e., non-differentially methylated) regardless of whether its gene is in either E-, E0 or E+. However, if a gene is differentially expressed i.e., in clusters E- or  E+, an associated CpG site, if also differentially methylated, has highest probability of being hypermethylated, i.e. in M+ (with probabilities $0.41$ and $0.24$, respectively).

To illustrate the dependence between gene and CpG cluster memberships, panel C in Figure \ref{idiffomix_fig:real_data_result} details the conditional probabilities of a gene belonging to cluster $k$ given a single CpG site associated with the gene belongs to cluster $l$, computed using Bayes' theorem, given $\hat{\bm{\tau}}$ and $\hat{\bm{\pi}}$. That is, 
$$ {\mathbb P}\{\mbox{Gene in cluster }k | \mbox{CpG in cluster }l\} = \frac{\hat\tau_{k}\hat\pi_{l|k}}{\sum_{k'=1}^{K}\hat\tau_{k'}\hat\pi_{l|k'}}.$$

Here, a gene has highest probability of being in cluster E0 (i.e., non-differentially expressed) regardless of whether its CpG site is in M-, M0 or M+. Thereafter, if a CpG site is in M-, its associated gene, if also differentially expressed, has highest probability of being in E+ (with probability $0.28$), while if a CpG site is in M+, its associated gene, if also differentially expressed, has highest probability of being E- (with probability $0.43$).

For comparison purposes, as the log-fold changes and differences in \textit{M}-values are approximately Gaussian distributed, \texttt{mclust} is fitted to each data type on chromosome 7 independently. 
Out of the $772$ genes located on chromosome 7, \texttt{idiffomix} identified $30$ genes as DEGs and $22$ genes as non-DEGs that were not detected as such by \texttt{mclust}.
For the $4,790$ CpG sites associated with promoter regions of genes on chromosome 7, \texttt{idiffomix} identified $36$ CpG sites as DMCs and $41$ CpG sites as non-DMCs that were not detected as such by \texttt{mclust}.

\subsubsection{Genes of interest}
\begin{figure}[H]
    \centering
    \includegraphics[width=\textwidth,height=9cm]{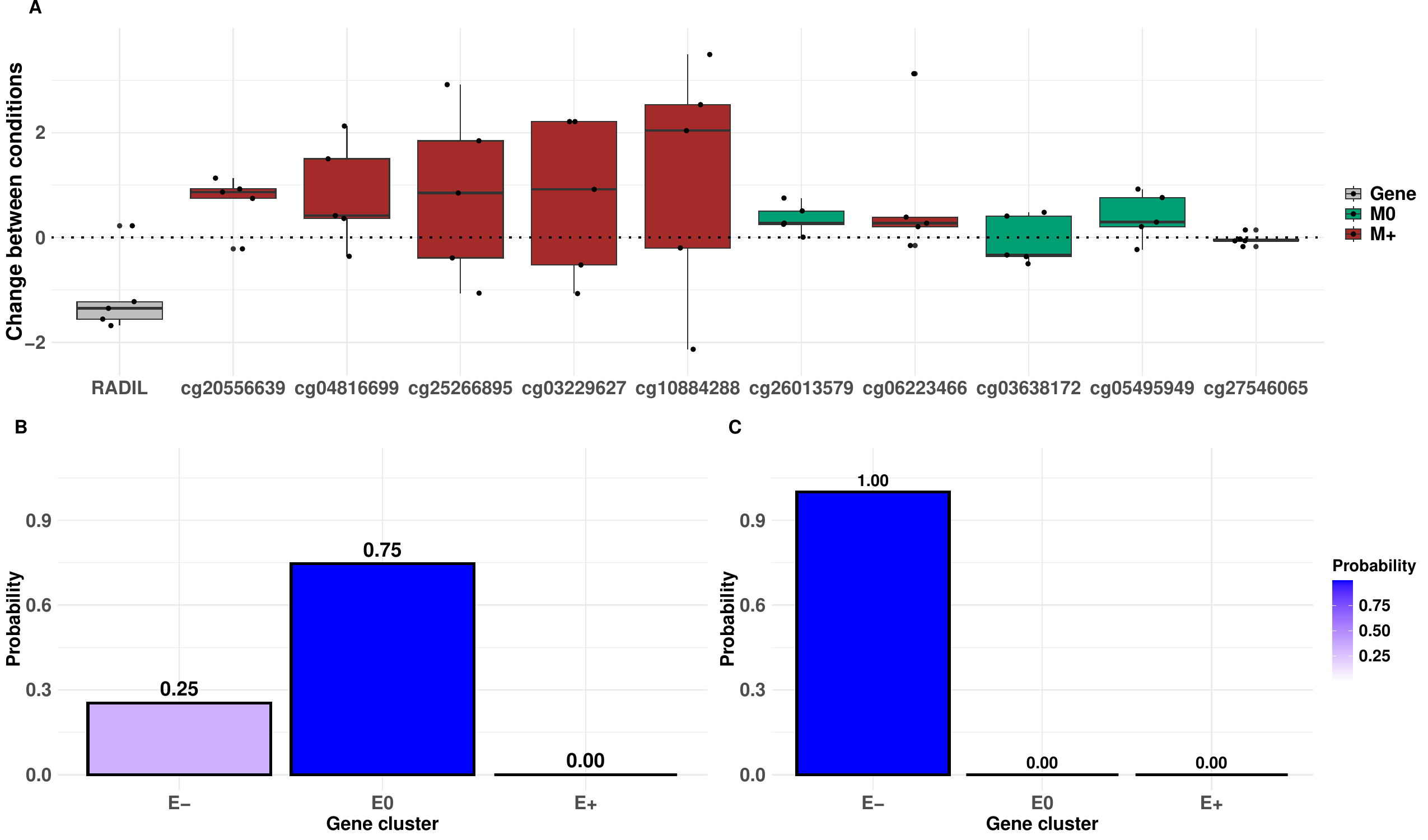}
    \caption{\textbf{Comparison of results for independent and integrated analyses for \textit{RADIL} on chromosome 7}\\ \small
    (A) Log-fold change in gene expression levels (grey) and differences in \textit{M}-values between tumour and normal samples, coloured by inferred idiffomix cluster (hypermethylated CpG sites, M+ in brown; non-differentially methylated CpG sites, M0 in green); (B) posterior probability of \textit{RADIL} belonging to the E-, E0 and E+ clusters under \texttt{mclust}, (C) posterior probability of \textit{RADIL} belonging to the E-, E0 and E+ clusters when jointly modelled with methylation data under \texttt{idiffomix}. Larger posterior probabilities are represented by increasingly dark shades of blue.}
    \label{idiffomix_fig:RADIL}
\end{figure}

Genes for which the inferred differential expression status differed between the independent and integrated analyses are of particular interest. When the two data types are modelled jointly, $6$ of the $10$ CpG sites linked to the \textit{RADIL} gene were deemed to be DMCs in the M+ cluster and $4$ were inferred to belong to the M0 cluster (Figure \ref{idiffomix_fig:RADIL}, panel A). Interestingly, when the RNA-Seq data were modelled independently the \textit{RADIL} gene, with a median log-fold change below $-1$, was identified to be non-differentially expressed, as illustrated in panel B of Figure \ref{idiffomix_fig:RADIL}. However, modelling the RNA-Seq data and the methylation data together under \texttt{idiffomix} suggested \textit{RADIL} was a DEG, belonging to the E- cluster, as illustrated in panel C of Figure \ref{idiffomix_fig:RADIL}. 

\begin{figure}[H]
    \centering
    \includegraphics[width=\textwidth]{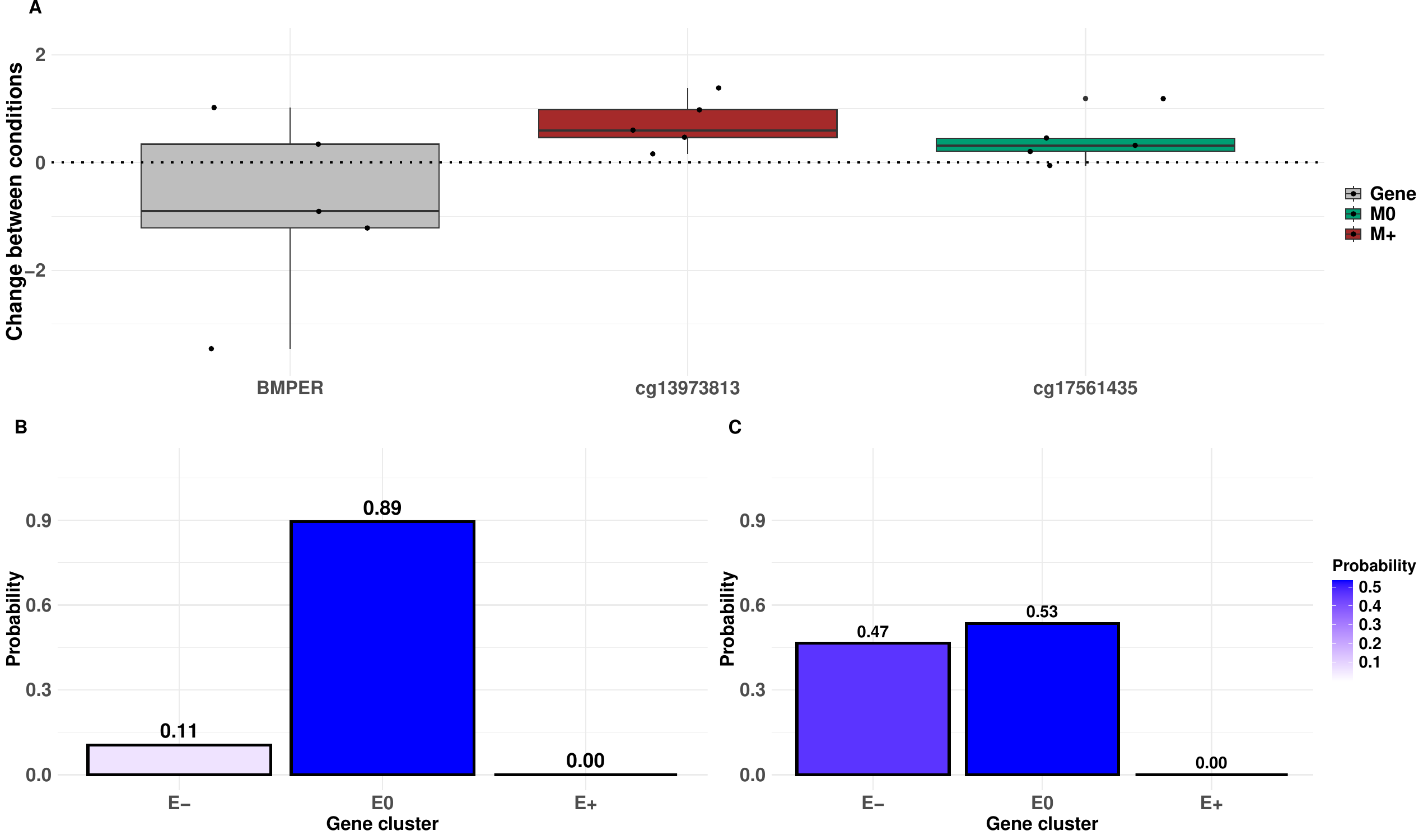}
    \caption{\textbf{Comparison of results for independent and integrated analyses for \textit{BMPER} on chromosome 7.} \\ \small
    (A) Log-fold change in gene expression levels (grey) and differences in \textit{M}-values between tumour and normal samples, coloured by inferred idiffomix cluster (hypermethylated CpG sites, M+ in brown; non-differentially methylated CpG sites, M0 in green); (B) posterior probability of \textit{BMPER} belonging to the E-, E0 and E+ clusters under \texttt{mclust} (C) posterior probability of \textit{BMPER} belonging to the E-, E0 and E+ clusters when jointly modelled with methylation data under \texttt{idiffomix}. Larger posterior probabilities are represented by increasingly dark shades of blue.}
    \label{idiffomix_fig:BMPER}
\end{figure}

Another aspect of interest are the gene's clustering uncertainties. Figure \ref{idiffomix_fig:BMPER} shows, in panel A, the log-fold change and difference in \textit{M}-values for \textit{BMPER} and its associated CpG sites while panels B and C show the posterior probabilities of cluster membership for \textit{BMPER} under \texttt{mclust} and \texttt{idiffomix} respectively. The resulting clustering uncertainty for \textit{BMPER} is higher under \texttt{idiffomix} than under \texttt{mclust} as, while the gene expression levels (panel A) suggest the gene is likely to be non-differential, of the two CpG sites linked to the gene, one is inferred to be hypermethylated and the other as non-differential. Consequently, the posterior probability of \textit{BMPER} belonging to the non-differential E0 cluster under \texttt{idiffomix} is smaller than that under \texttt{mclust}, while the probability of \textit{BMPER} being a DEG in the E- cluster is larger. 
Thus, while the MAP cluster for \textit{BMPER} is E0 under both independent and integrated approaches, the clustering uncertainty is intuitively higher under \texttt{idiffomix}, given the observed data. Clustering uncertainties for other genes of interest are provided in Appendix A.5.

\begin{figure}[H]
    \centering
    \includegraphics[width=\textwidth]{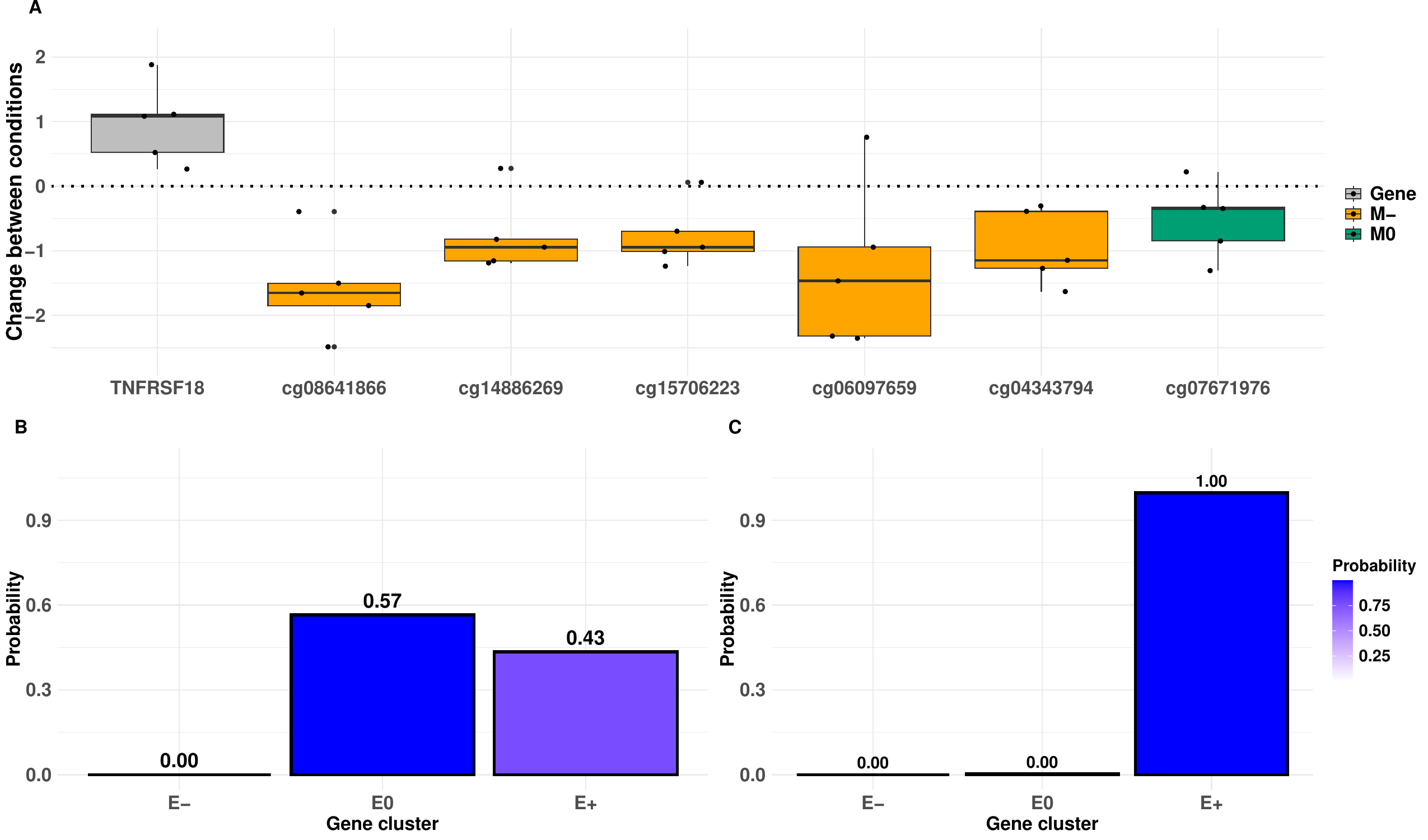}
    \caption{\textbf{Comparison of results for independent and integrated analyses \textit{TNFRSF18} on chromosome 1}\\ \small
    (A) Log-fold change in gene expression levels (grey) and differences in \textit{M}-values between tumour and normal samples, coloured by inferred idiffomix cluster (hypomethylated CpG sites, M- in yellow; non-differentially methylated CpG sites, M0 are green); (B) posterior probability of \textit{TNFRSF18} belonging to the E-, E0 and E+ clusters under \texttt{mclust} (C) posterior probability of \textit{TNFRSF18} belonging to the E-, E0 and E+ clusters when jointly modelled with methylation data under \texttt{idiffomix}. Larger posterior probabilities are represented by increasingly dark shades of blue.}
    \label{idiffomix_fig:tnfrsf18}
\end{figure}

\textit{TNFRSF18}, \textit{GPX7} and \textit{RAD51}
are of particular interest as they are known to play key roles in the development and progression of breast cancer \citep{wiegmans2014rad51,rusolo2017comparison,xiong2019tumor}.
Figure \ref{idiffomix_fig:tnfrsf18} (panel A) shows that the median log-fold change for \textit{TNFRSF18}
is $>$1, and that of the $6$ CpG sites linked to the gene, $5$ are DMCs in M- and $1$ is non-differential in M0 when the two data types are modelled jointly. When the expression data of \textit{TNFRSF18} are modelled independently under \texttt{mclust} (Figure \ref{idiffomix_fig:tnfrsf18} panel B), \textit{TNFRSF18} has posterior probabilities of $0.57$ and $0.43$ of being in clusters E0 and E+ respectively. However, when the expression and methylation data are modelled jointly using \texttt{idiffomix} (Figure \ref{idiffomix_fig:tnfrsf18} panel C), the posterior probability of  \textit{TNFRSF18} being upregulated in cluster E+ is $1$. 
These results show that when the RNA-Seq and methylation data are jointly modelled, different insights are revealed than when the two data types are modelled independently. 
Posterior probability cluster membership distributions, and their corresponding insights, for \textit{GPX7} and \textit{RAD51} are available in Appendix A.6.

\begin{figure}[H]
    \centering
    \includegraphics[width=\linewidth]{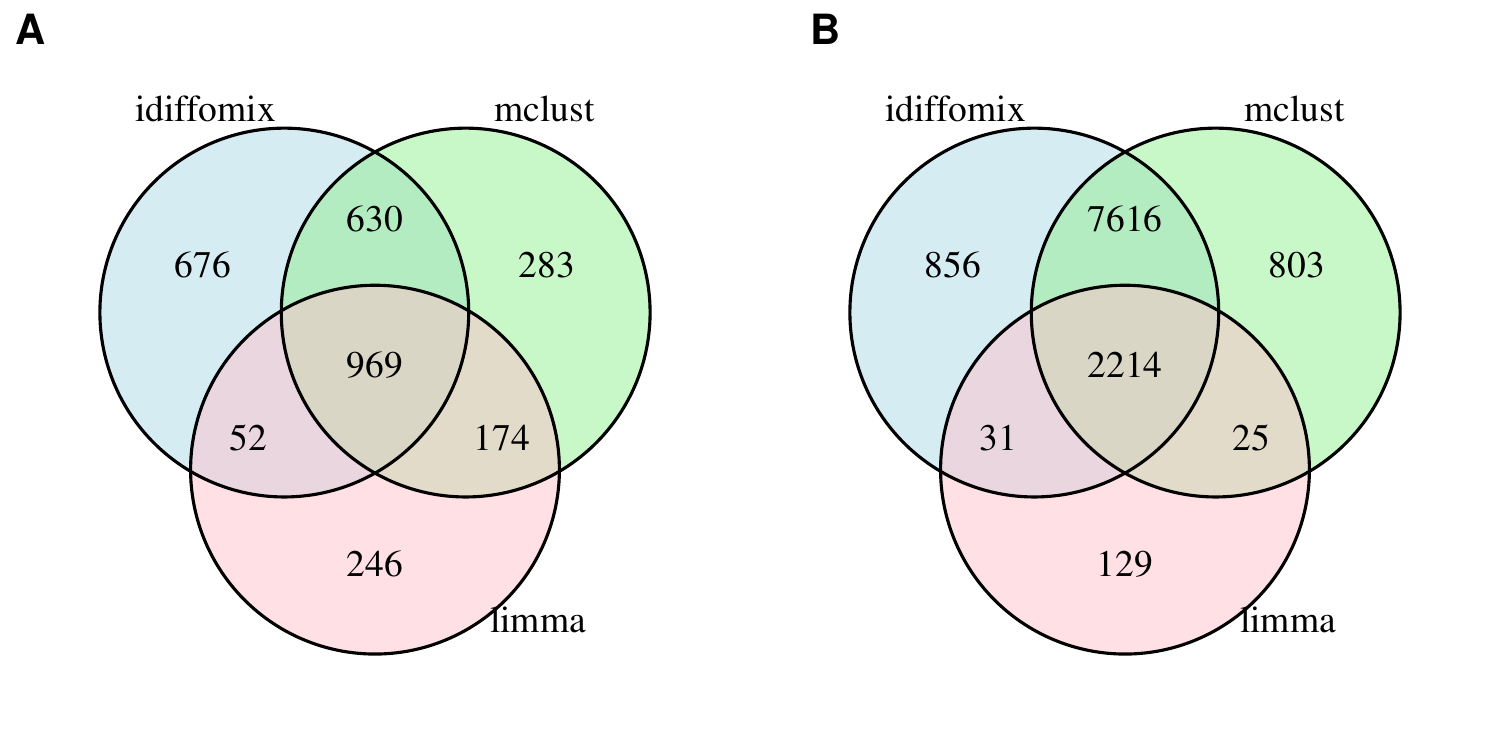}
    \caption{\textbf{Comparison of results for DEGs and DMCs between independent and integrated analyses.} \\ \small Venn diagrams showing the intersection of (A) DEGs and (B) DMCs identified under  \texttt{idiffomix}, \texttt{mclust}, and \texttt{limma}.}
    \label{idiffomix_fig:venn_dmc_deg}
\end{figure}

The two data types were also modelled independently using \texttt{limma} for comparison purposes. In contrast to \texttt{idiffomix} which identified $2,327$ DEGs and $10,717$ DMCs, \texttt{mclust} identified $2,056$ DEGs and $10,658$ DMCs while \texttt{limma} identified $1,441$ DEGs and $2,399$ DMCs. 
Figure \ref{idiffomix_fig:venn_dmc_deg} shows Venn diagrams illustrating the intersection of DEGs and DMCs, in panels A and B respectively,  identified under \texttt{idiffomix}, \texttt{mclust}, and \texttt{limma}, highlighting their common and method-specific findings.
Out of the $5$ genes of interest discussed here, \texttt{idiffomix} identified $3$ to be DEGs, while \texttt{mclust} identified \textit{RAD51} as a DEG, and under \texttt{limma} only \textit{TNFRSF18} was identified to be a DEG.  

\subsubsection{Gene enrichment analysis}
A gene enrichment analysis was performed for the DEGs and DMCs identified under the joint \texttt{idiffomix} model, and under the independent \texttt{mclust} and \texttt{limma} approaches using Gene Ontology (GO) and Kyoto Encyclopedia of Genes and Genomes (KEGG) terms \citep{KEGG, GO}. The GO enrichment analysis of \texttt{idiffomix} DEGs indicated they were associated with $1,299$ significant biological pathways, while DEGs under \texttt{mclust} and \texttt{limma} applied to the RNA-Seq data identified $1,442$ and $647$ significant biological pathways respectively. Similarly, the GO analysis of the \texttt{idiffomix} DMCs unveiled
$457$ significant biological pathways whereas \texttt{mclust} and \texttt{limma} identified $414$ and $83$ respectively.  Several biological processes identified under \texttt{idiffomix}, but not under the other two methods, such as \textit{MAPK cascade}, \textit{ERK1 and ERK2 cascade}, \textit{extracellular matrix organization} and \textit{BMP signaling pathway} play an essential role in breast cancer development and prognosis \citep{MAPK,BMP,EMO}.
The KEGG analysis of \texttt{idiffomix} DEGs revealed associations with $27$ significant pathways (adjusted $p$-value $<0.05$), while \texttt{mclust} and \texttt{limma} DEGs identified $32$ and $13$ significant pathways, respectively. Similarly, the KEGG analysis mapped \texttt{idiffomix} DMCs to genes associated with $17$ significant metabolic and signalling pathways, while DMCs under \texttt{mclust} and \texttt{limma} were mapped to genes associated with $18$ and $4$ significant pathways respectively.  Several pathways associated with \texttt{idiffomix} DEGs, but not identified under the other two methods, such as the \textit{cAMP signaling pathway}, \textit{ECM-receptor interaction}, \textit{Hippo signaling pathway} and \textit{Cell adhesion molecules} have also been shown to play key roles in breast cancer development \citep{ Hippo, adhesion,camp}. The top 10 biological processes and pathways associated with the \texttt{idiffomix} DEGs and DMCs 
are presented in Appendix A.7.

\section{Discussion}
While there are inherent, biological dependencies between gene sequencing and methylation array data, analyses that aim to identify differential gene expression and methylation typically do so via independent analyses of both data types. 
Here, a joint mixture model approach \texttt{idiffomix} is proposed that integrates both data types at the modelling stage by directly modelling the nested structure of CpG sites in gene promoter regions. 
The method does not assume a fixed pattern of associations between expression and promoter methylation (e.g., global anti-correlation for all genes/CpGs); rather, it learns global patterns from the data. This allows for a  genome-wide, cross-omics analysis that simultaneously identifies DMCs and DEGs.  
Simulation studies and application to data from a breast cancer study demonstrated the benefit of integrating both data types at the modelling stage, providing a joint analysis.

While log-fold changes of transformed RNA-Seq and differences in \textit{beta} valued methylation data were modelled here, allowing for a joint Gaussian mixture model, such transformations can make results less  biologically interpretable. Relaxing the Gaussian assumption and employing distributions that 
model the inherent data distributions directly could improve model performance and interpretability.

Information on the complete set of proteins that is expressed by a cell or tissue under specific conditions is captured in proteomics data, which can be obtained using high-throughput technologies like mass spectroscopy. Integrating proteomics with other omics data, such as gene expression data, has been shown to capture the complexity of biological systems and provide deeper insights into gene regulation and cellular functions \citep{kumar2016integrating}. The joint mixture model approach could be enhanced to jointly model proteomics data along with gene expression and DNA methylation data.

It is well established that methylation patterns and gene expression regulation are not only dependent on one another but also on several other factors including environmental stress, food habits, etc \citep{muralidharan2022environmental}. The \texttt{idiffomix} method could be further developed to facilitate modelling of the influence on the two data types of such environmental factors. For example, employing a mixtures of experts approach \citep{Claire}, where cluster membership probabilities are modelled as dependent on concomitant factors, would allow for such covariates to be incorporated within \texttt{idiffomix}'s joint mixture model framework.

\section{Conclusions}
The development and application of the \texttt{idiffomix} joint mixture model for integrated differential analysis of multi-omics data illustrated the advantages of integrating RNA-Seq and methylation data when identifying DEGs and DMCs, over analysing such data separately.  The \texttt{idiffomix} model provides insight to the complex associations between gene expression and methylation changes, leading to a deeper understanding of the transcriptional and epigenetic landscape. Future work on this model will aim to refine and expand its capabilities, building on our current findings.




\section{\textbf{Acknowledgements}}
This publication has emanated from research conducted with the financial support of Research Ireland under grant number 18/CRT/6049. For the purpose of Open Access, the author has applied a CC BY public copyright licence to any Author Accepted Manuscript version arising from this submission. For the purpose of Open Access, the author has applied a CC BY public copyright licence to any Author Accepted Manuscript version arising from this submission. 



\bibliographystyle{plainnat}
\bibliography{Main.bib}{}   

\appendix

\section{Appendix to \lq Integrated differential analysis of multi-omics data using a joint mixture model: \texttt{idiffomix}\rq  \space by Majumdar et al.}
\subsection{EM algorithm for joint mixture model}
The complete data log-likelihood function for the joint mixture model is,
\begin{equation}
\begin{aligned}
    \ell_C(\bm{\tau},\bm{\pi}, \bm{\theta}, \bm{\phi}| \bm{U}, \bm{V},{\bm{X}}, {\bm{Y}}) = & \sum\limits_{g=1}^G \sum\limits_{k=1}^K  \sum\limits_{n=1}^N u_{gk}\log p({x}_{gn}|\bm{\theta}_k) \\ & +  \sum\limits_{g=1}^G\sum\limits_{c=1}^{C_g} \sum\limits_{l=1}^L \sum\limits_{n=1}^N  \; v_{gcl}\log p({y}_{gcn}|\bm{\phi}_{l})\\
    & + \sum\limits_{g=1}^G \sum\limits_{k=1}^K u_{gk}\log \tau_k + \sum_{g=1}^{G}\sum_{k=1}^{K}\sum\limits_{c=1}^{C_g} \sum\limits_{l=1}^L u_{gk}v_{gcl}\log\pi_{l|k}
    \end{aligned}
    \label{idiffomix_appeqn:complete_llk}
\end{equation}    
\paragraph{M-step of EM algorithm for the joint mixture model}
The complete data log-likelihood function in (\ref{idiffomix_appeqn:complete_llk}) is maximised w.r.t. the parameters ${\tau}_k$, ${\pi}_{l|k}$, $\bm{\theta}_{k} = $ (${\mu}_{k}$, ${\sigma}^2$), and $\bm{\phi}_{l} = $ (${\lambda}_{l}$, ${\rho}^2$) to calculate $\hat\tau_k$, $\hat\pi_{l|k}$, $\hat{{\sigma}}^2$ and $\hat{{\rho}}^2$. For parsimony, the standard deviations are constrained to be equal across clusters and are calculated as the weighted average of the individual standard deviations, such that $\hat\sigma^2 = \sum_{k=1}^K \tau_k  \sigma_{k}^2$ and $\hat\rho^2 = \sum_{l=1}^L (\sum_{k=1}^K \pi_{l|k} \tau_k)  \sigma_{k}^2$. The other maximised parameter estimates are then:
\begin{align*}
    \hat\tau_k &= \frac{\sum_{g=1}^{G}u_{gk}}{G}, &
    \hat\pi_{l|k} &= \frac{\sum_{g=1}^{G}\sum_{c=1}^{C_g}u_{gk}v_{gcl}}{\sum_{g=1}^{G}  u_{gk}C_g}, \\
    \hat{{\mu}}_{k} &= \frac{\sum\limits_{g=1}^G \sum\limits_{n=1}^N u_{gk}\ {{x}}_{gn}}{N \sum\limits_{g=1}^G u_{gk}}, &
    \hat{{\sigma}}_{k}^2 &= \frac{\sum\limits_{g=1}^G \sum\limits_{n=1}^N u_{gk} ({{x}}_{gn}-\hat{{\mu}}_{k})^2}{N \sum\limits_{g=1}^G u_{gk}}, \\
    \hat{{\lambda}}_{l} &= \frac{\sum\limits_{g=1}^G  \sum\limits_{c=1}^{C_g} \sum\limits_{n=1}^N v_{gcl}\ {{y}}_{gcn}}{N \sum\limits_{g=1}^G \sum\limits_{c=1}^{C_g} v_{gcl}}, &
    \hat{{\rho}}_{l}^2 &= \frac{\sum\limits_{g=1}^G \sum\limits_{c=1}^{C_g} \sum\limits_{n=1}^N v_{gcl} ({{y}}_{gcn}-\hat{{\lambda}}_{l})^2}{N \sum\limits_{g=1}^G \sum\limits_{c=1}^{C_g} v_{gcl}}.
\end{align*}









\paragraph{E-step of EM algorithm for the joint mixture model}
In the E-step, the conditional expected values of $u_{gk}$, $v_{gcl}$ and $u_{gk}v_{gcl}$ are calculated given the observed data ${\bm{X}}$, ${\bm{Y}}$ and the estimated model parameters $\tau_{k}^{(t)}$, $\pi_{l|k}^{(t)}$, $\bm{\theta}_{k}^{(t)}$, and $\bm{\phi}_{l}^{(t)}$, at iteration $t$ i.e., we have to compute 
$\mathbb{E}(u_{gk}| {\bm{x}}_g, \tau_k, \bm{\pi}_{l|k}, \bm{\theta}_k )$, $\mathbb{E}(v_{gcl}| {\bm{y}}_{gc}, \bm{\pi}_{l|k}, \bm{\phi}_l)$, and $\mathbb{E}(u_{gk}v_{gcl}| {\bm{x}}_g, {\bm{y}}_{gc}, \tau_k, \bm{\pi}_{l|k}, \bm{\theta}_k, \bm{\phi}_l )$.
However, these expected values are intractable.

Here, instead, the following expected values are considered and can be derived from the complete data log-likelihood function (\ref{idiffomix_appeqn:complete_llk}):
\begin{equation}
\begin{aligned}
    \mathbb{E}(u_{gk}| {\bm{x}}_{g}, v_{g11}, v_{g12}, \cdots, v_{gC_gL}, \tau_k, \bm{\theta}_k, \pi_{1|k}, \cdots, \pi_{L|k})=  \frac{\tau_k \prod\limits_{n=1}^N p({x}_{gn}|\bm{\theta}_k) \prod\limits_{c=1}^{C_g} \prod\limits_{l=1}^L (\pi_{l|k})^{v_{gcl}}}{\sum\limits_{k=1}^K \{\tau_k \prod\limits_{n=1}^N p({x}_{gn}|\bm{\theta}_k) \prod\limits_{c=1}^{C_g} \prod\limits_{l=1}^L (\pi_{l|k})^{v_{gcl}} \}} , 
        \label{idiffomix_appeqn:e_u_gk}
    \end{aligned}
\end{equation}
$\mbox{ for } g = 1, 2, \cdots, G \mbox{, and } k = 1, 2, \cdots ,K,$ and
\begin{equation}
    \mathbb{E}(v_{gcl}|{\bm{y}}_{gc}, u_{g1},u_{g2},\cdots,u_{gK}, \bm{\phi}_l, \pi_{l|1}, \cdots, \pi_{l|K})=\frac{\prod\limits_{n=1}^N p({y}_{gcn}|\bm{\phi}_l) \prod\limits_{k=1}^K (\pi_{l|k})^{u_{gk}}}{\sum\limits_{l=1}^L \{\prod\limits_{n=1}^N p({y}_{gcn}|\bm{\phi}_l) \prod\limits_{k=1}^K (\pi_{l|k})^{u_{gk}}  \} } ,
    \label{idiffomix_appeqn:e_v_gcl}
\end{equation}
$\mbox{ for } g = 1, 2, \cdots, G,$ $ c=1, 2, \cdots, C_g$, and 
$ l = 1, 2, \cdots, L$. Given the conditional expected values in (\ref{idiffomix_appeqn:e_u_gk}) and (\ref{idiffomix_appeqn:e_v_gcl}) then:
\begin{equation}
\begin{aligned}
    \mathbb{P}(u_{gk}=1|{\bm{x}}_{g}, v_{g11}, v_{g12}, \cdots, & v_{gC_gL}, \tau_k, \bm{\theta}_k, \pi_{1|k}, \cdots, \pi_{L|k})= \\ & \frac{\tau_k \prod\limits_{n=1}^N p({x}_{gn}|\bm{\theta}_k) \prod\limits_{c=1}^{C_g} \prod\limits_{l=1}^L (\pi_{l|k})^{v_{gcl}}}{\sum\limits_{k=1}^K \{\tau_k \prod\limits_{n=1}^N p({x}_{gn}|\bm{\theta}_k) \prod\limits_{c=1}^{C_g} \prod\limits_{l=1}^L (\pi_{l|k})^{v_{gcl}} \}} , 
        \label{idiffomix_appeqn:p_u_gk}
    \end{aligned}
\end{equation}

\begin{equation}
\begin{aligned}
    \mathbb{P}(v_{gcl}=1|{\bm{y}}_{gc}, u_{g1},u_{g2},\cdots, & u_{gK}, \bm{\phi}_l, \pi_{l|1}, \cdots, \pi_{l|K})= \\ & \frac{\prod\limits_{n=1}^N p({y}_{gcn}|\bm{\phi}_l) \prod\limits_{k=1}^K (\pi_{l|k})^{u_{gk}}}{\sum\limits_{l=1}^L \{\prod\limits_{n=1}^N p({y}_{gcn}|\bm{\phi}_l) \prod\limits_{k=1}^K (\pi_{l|k})^{u_{gk}}  \} }, 
    \label{idiffomix_appeqn:p_v_gcl}
        \end{aligned}
\end{equation}
$ \mbox{ for } g = 1, 2, \cdots, G, k = 1, 2, \ldots, K, c=1, 2, \cdots, C_g$, and 
$ l = 1, 2, \cdots, L$.

The probabilities given in (\ref{idiffomix_appeqn:p_u_gk}) and (\ref{idiffomix_appeqn:p_v_gcl}) could be used to sample values of $\bm{U}$ and $\bm{V}$ using an MCMC algorithm. However, applying an MCMC algorithm to the large dataset considered here where $G = 15,722$ and $C = 94,873$ would be computationally expensive. Therefore, for computational efficiency, an algorithm similar to that employed in \cite{BMMSTcoordinateascent} and \cite{coordinateascent} is used to compute the required conditional expected values. Under this approach, the expected values given in (\ref{idiffomix_appeqn:e_u_gk}) and (\ref{idiffomix_appeqn:e_v_gcl}) are iteratively computed until convergence (after $S$ iterations). The computed values are then used at convergence to calculate the required expected values: 
\begin{align*}
    \mathbb{E}(u_{gk}| \cdots) \approx  &  u_{gk}^{(S)} = \hat{u}_{gk},  \hspace{1cm}
    \mathbb{E}(v_{gcl}| \cdots ) \approx   v_{gcl}^{(S)} = \hat{v}_{gcl}, \hspace{0.2cm} \text{and} \\
    &\mathbb{E}(u_{gk}v_{gcl}| \cdots ) \approx  u_{gk}^{(S)}v_{gcl}^{(S)} = \widehat{{u}_{gk}v_{gcl}}.
\end{align*}
In practice, here $S \approx 10$ iterations were required to achieve convergence per iteration of the EM algorithm. 



\clearpage
\subsection{Performance metrics for simulated datasets}
\begin{table}[htb]
    \centering
        \caption{Mean performance metrics (standard deviations in parentheses) for 100 simulated datasets  given $\bm{\pi}$ under case 3 from Table 1.}
        \begin{tabular}{c}
        \begin{subtable}[t]{\textwidth}
            \centering
            \captionsetup{labelformat=empty}
                        \caption{(a) DEG identification performance}
            \begin{tabular}{c c c c c }
                \hline
                & FDR & Sensitivity & Specificity & ARI \\
                \hline
                \texttt{idiffomix} & \textbf{0.036} (0.025) & 0.771 (0.072) & \textbf{0.993} (0.006) & 0.766 (0.061) \\ \hline
               \texttt{mclust} & 0.102 (0.049) & \textbf{0.873} (0.046) & 0.975 (0.015) & \textbf{0.800} (0.041) \\ \hline
                \texttt{limma} & 0.038 (0.021) & 0.764 (0.064) & \textbf{0.993} (0.005) & 0.760 (0.059) \\ \hline
            \end{tabular}
        \end{subtable} \\
        \vspace{0.3cm} 
        \begin{subtable}[t]{\textwidth}
            \centering
            \captionsetup{labelformat=empty}
              \caption{(b) DMC identification performance}
            \begin{tabular}{c c c c c }
                \hline
                & FDR & Sensitivity & Specificity & ARI \\
                \hline
                \textbf{\texttt{idiffomix}} & \textbf{0.009} ( 0.003) & \textbf{1.000} ($<$ 0.001) & \textbf{0.994} (0.002) & \textbf{0.985} (0.004) \\ \hline
               \textbf{\texttt{mclust}} & \textbf{0.009} (0.003) & \textbf{1.000} ($<$ 0.001) & \textbf{0.994} (0.002) & \textbf{0.985} (0.004) \\ \hline
                \texttt{limma} & 0.046 (0.004) & 1.000 ($<$ 0.001) & 0.968 (0.003) & 0.924 (0.006) \\ \hline
            \end{tabular}
        \end{subtable}
    \end{tabular}
\end{table}

\begin{table}[htb]
    \centering
    \caption{Mean ARI values (with standard deviations in parentheses) for 100 simulated datasets  given $\bm{\pi}$ under case 3 from Table 1 for comparing clustering solutions between methods.}
    \begin{tabular}{c}
        \begin{subtable}[t]{0.7\textwidth}
            \centering
            \captionsetup{labelformat=empty}
            \caption{(a) ARI for DEG identification}
            \begin{tabular}{l c }
                \hline
                Method & Mean ARI \\
                \hline
                \texttt{idiffomix} vs \texttt{mclust} & 0.837 (0.100) \\
                \texttt{idiffomix} vs \texttt{limma} & 0.936 (0.041) \\
                \hline
            \end{tabular}
        \end{subtable} \\
        \vspace{0.3cm} 
        \begin{subtable}[t]{0.7\textwidth}
            \centering
            \captionsetup{labelformat=empty}
            \caption{(b) ARI for DMC identification}
            \begin{tabular}{l c }
                \hline
                Method & Mean ARI \\
                \hline
                \texttt{idiffomix} vs \texttt{mclust} & 0.999 (0.001) \\
                \texttt{idiffomix} vs \texttt{limma} & 0.915 (0.007) \\
                \hline
            \end{tabular}
        \end{subtable}
    \end{tabular}
\end{table}

\clearpage

\subsection{Time complexity}

 \begin{figure}[h!]
    \centering
    \includegraphics[width=\textwidth]{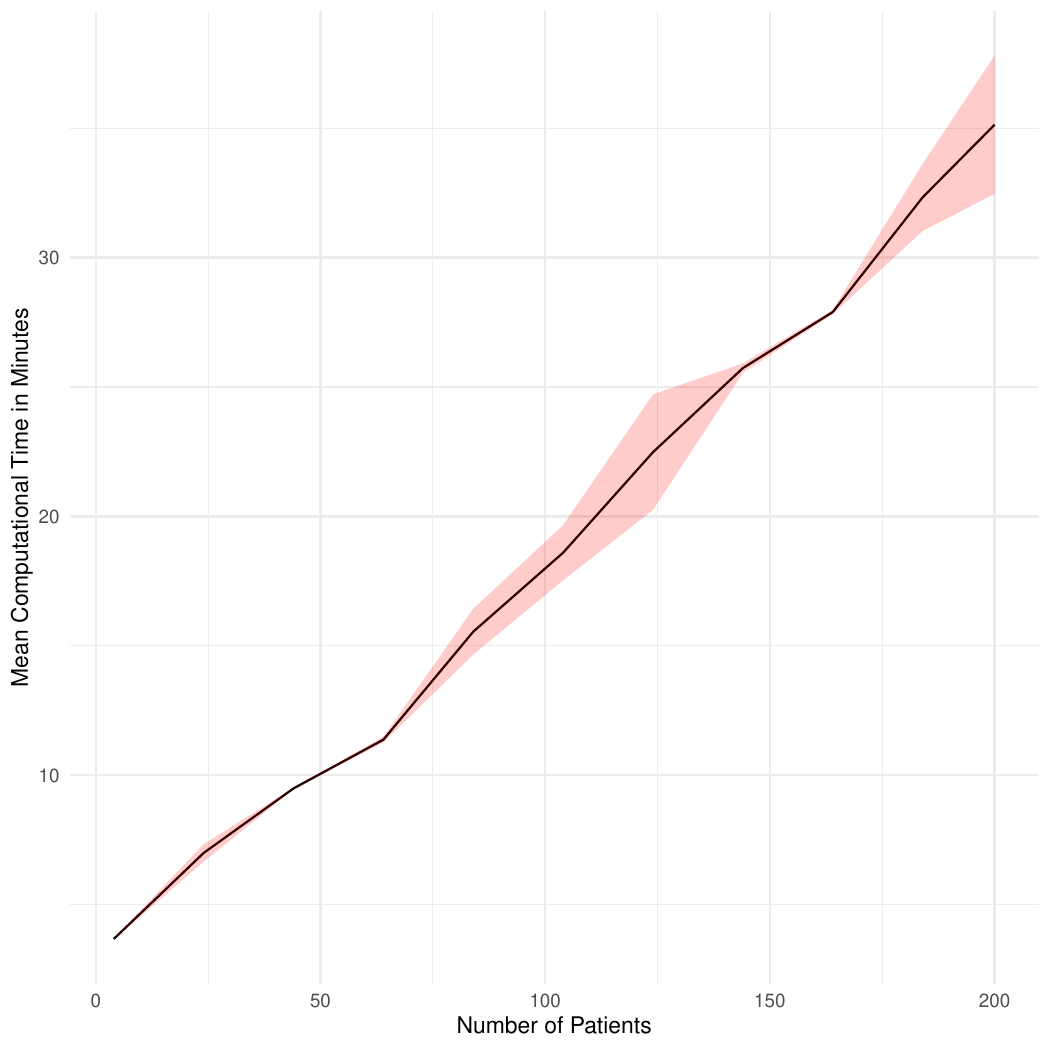}
    \caption{Mean computational time for fitting the \texttt{idiffomix} model, with 95\% confidence intervals, as
the number of patients, $N$, is increased from 4 to 200. As the complexity of the algorithm with respect to $N$ is proportional to $N$, as the number of patients increases the computational cost scales linearly.}
    \label{idiffomix_appfig:time_complexity}
\end{figure}

\clearpage

\subsection{Idiffomix results for all chromosomes from TCGA breast cancer data}
 \begin{figure}[htb]
    \centering
    \includegraphics[width=\textwidth]{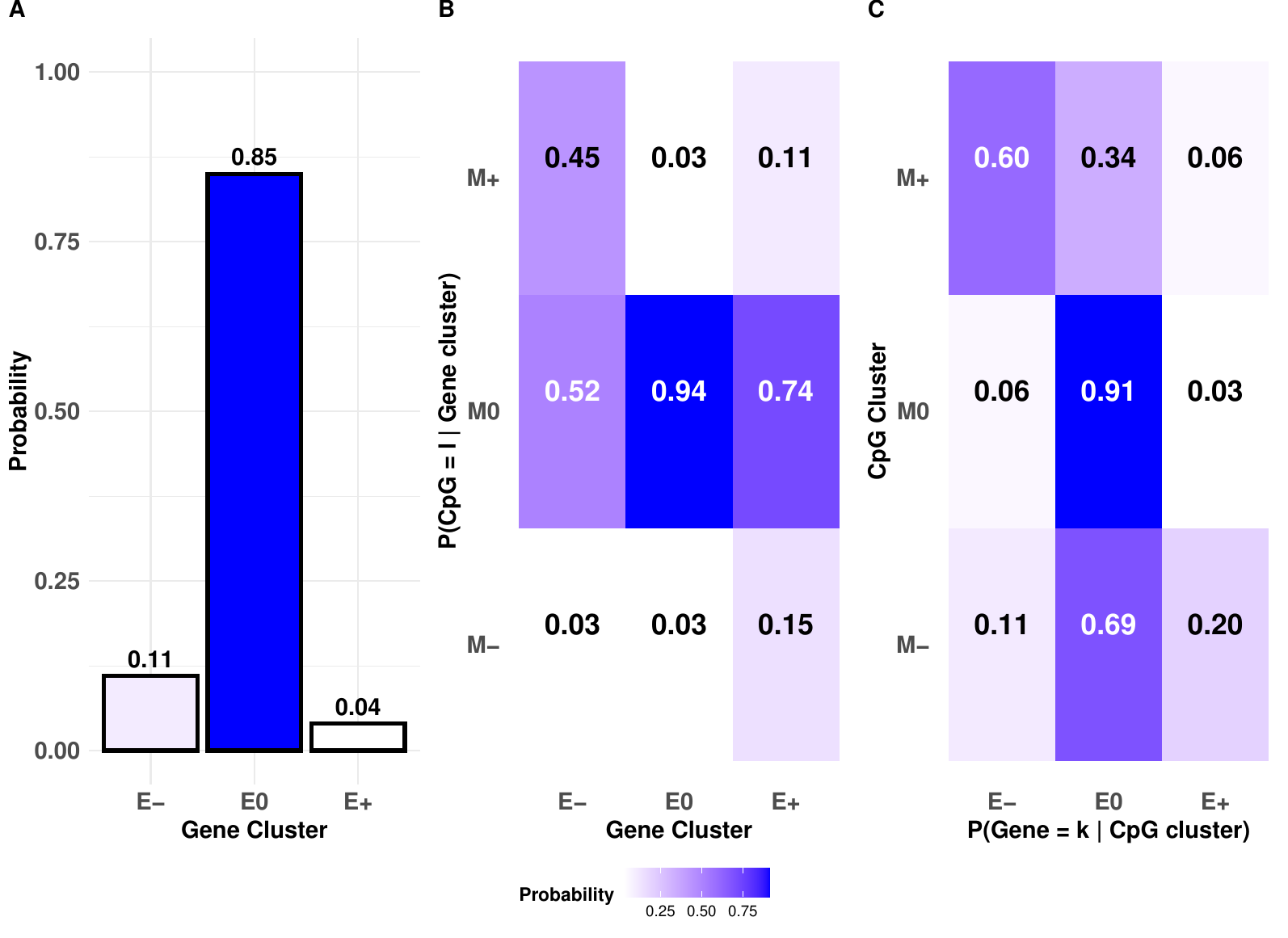}
    \caption{\texttt{Idiffomix} applied to chromosome 1 of TCGA breast cancer data: (A) estimated cluster  membership probabilities $\hat{\bm{\tau}}$, (B) the estimated matrix $\hat{\bm{\pi}}$ of conditional probabilities of CpG site methylation status given gene cluster membership, (c) conditional probabilities of genes belonging to cluster $k$ given a single CpG site associated with the gene belongs to cluster $l$.}
\end{figure}
 \begin{figure}[htb]
    \centering
    \includegraphics[width=\textwidth]{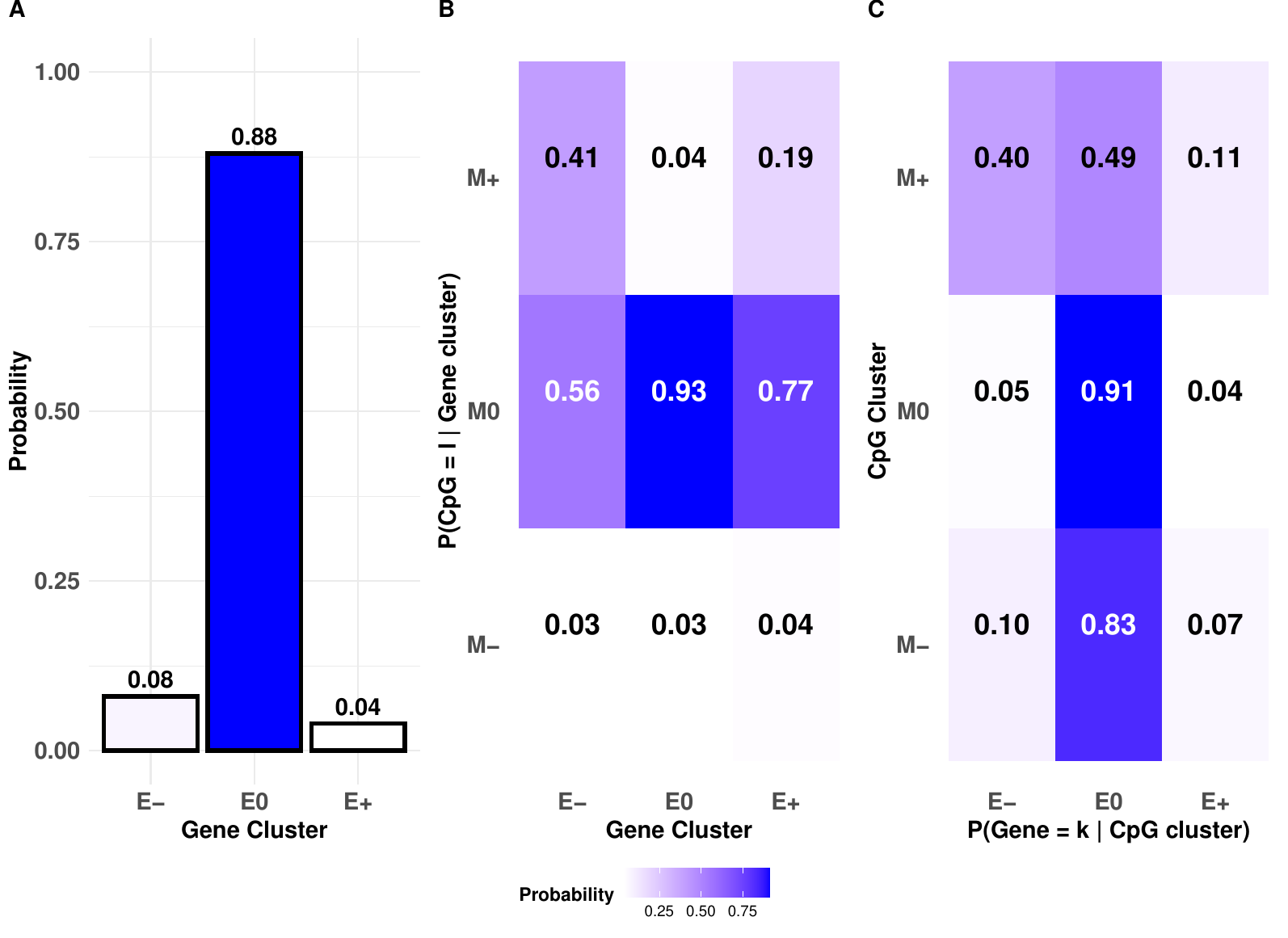}
    \caption{\texttt{Idiffomix} applied to chromosome 2 of TCGA breast cancer data: (A) estimated cluster  membership probabilities $\hat{\bm{\tau}}$, (B) the estimated matrix $\hat{\bm{\pi}}$ of conditional probabilities of CpG site methylation status given gene cluster membership, (c) conditional probabilities of genes belonging to cluster $k$ given a single CpG site associated with the gene belongs to cluster $l$.}
\end{figure}
 \begin{figure}[htb]
    \centering
    \includegraphics[width=\textwidth]{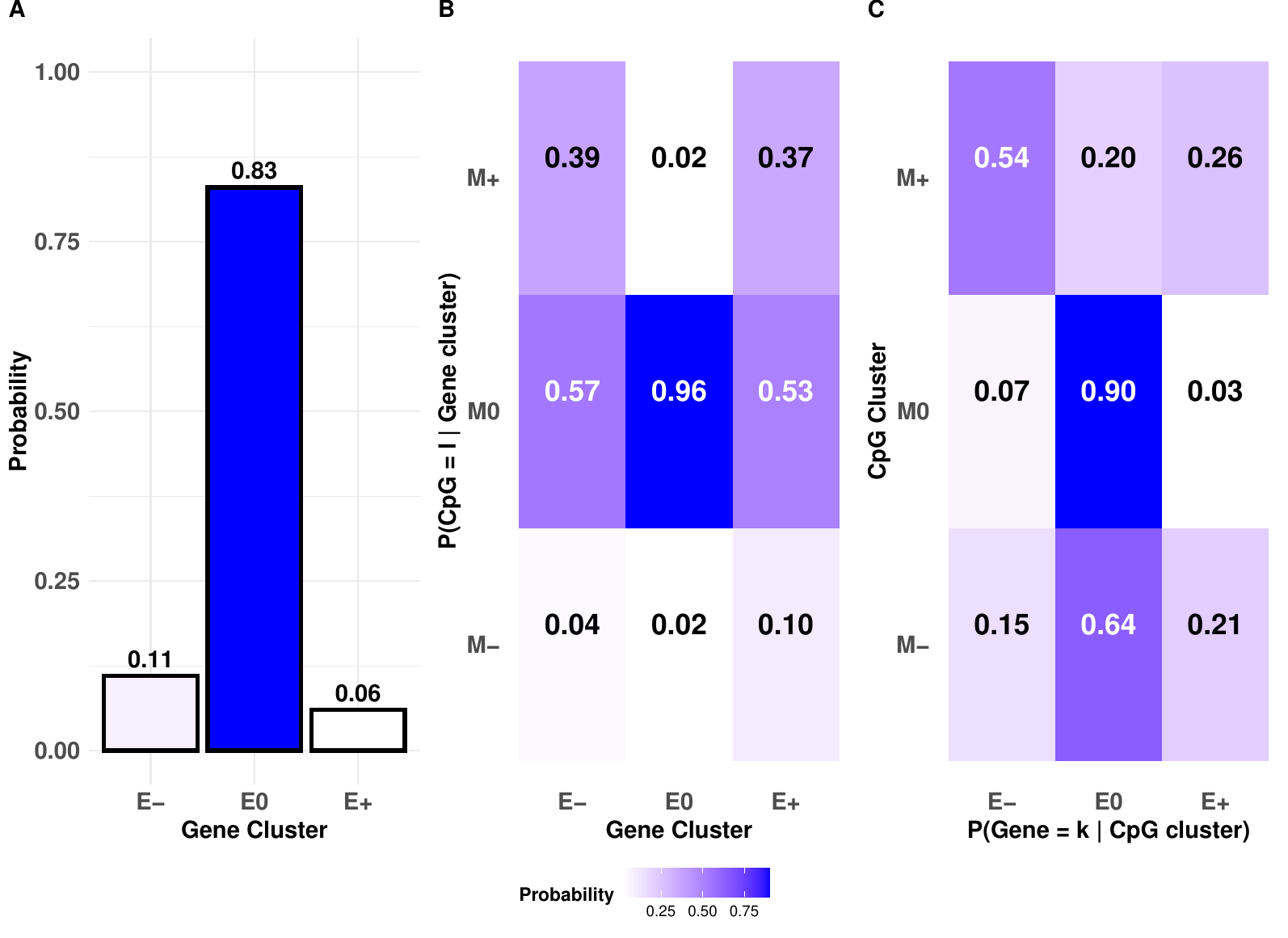}
    \caption{\texttt{Idiffomix} applied to chromosome 3 of TCGA breast cancer data: (A) estimated cluster  membership probabilities $\hat{\bm{\tau}}$, (B) the estimated matrix $\hat{\bm{\pi}}$ of conditional probabilities of CpG site methylation status given gene cluster membership, (c) conditional probabilities of genes belonging to cluster $k$ given a single CpG site associated with the gene belongs to cluster $l$.}
\end{figure}
 \begin{figure}[htb]
    \centering
    \includegraphics[width=\textwidth]{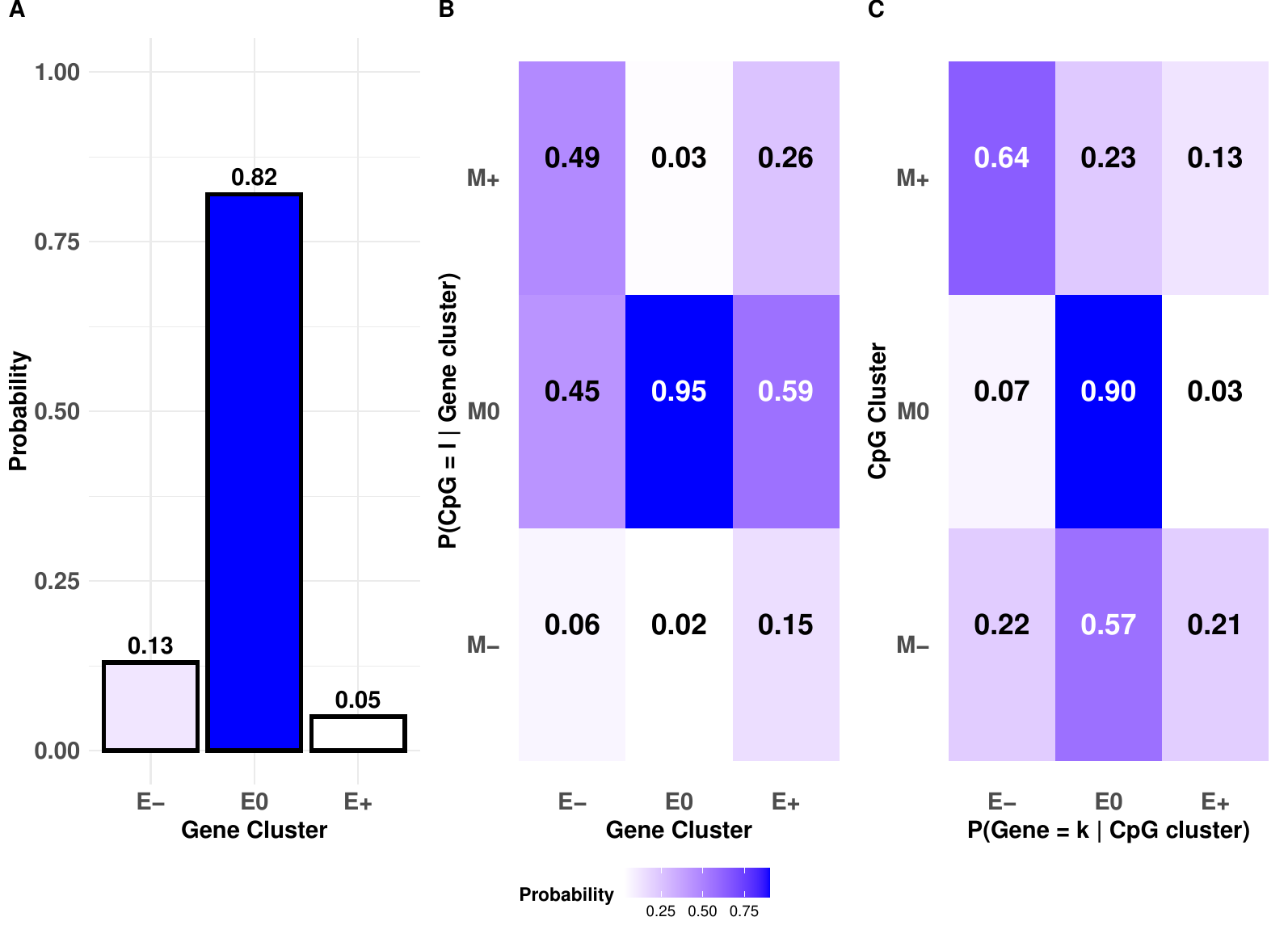}
    \caption{\texttt{Idiffomix} applied to chromosome 4 of TCGA breast cancer data: (A) estimated cluster  membership probabilities $\hat{\bm{\tau}}$, (B) the estimated matrix $\hat{\bm{\pi}}$ of conditional probabilities of CpG site methylation status given gene cluster membership, (c) conditional probabilities of genes belonging to cluster $k$ given a single CpG site associated with the gene belongs to cluster $l$.}
\end{figure}
 \begin{figure}[htb]
    \centering
    \includegraphics[width=\textwidth]{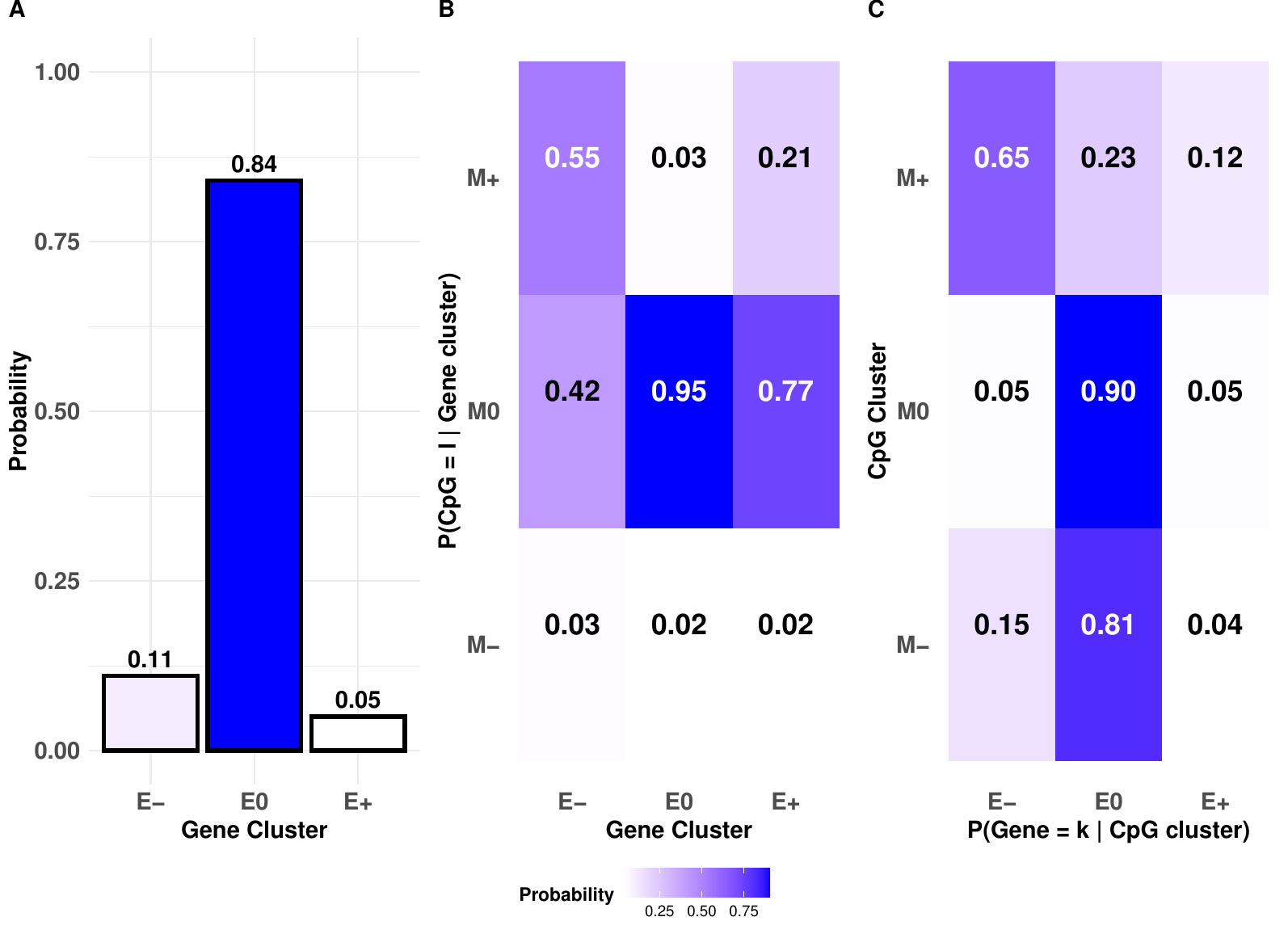}
    \caption{\texttt{Idiffomix} applied to chromosome 5 of TCGA breast cancer data: (A) estimated cluster  membership probabilities $\hat{\bm{\tau}}$, (B) the estimated matrix $\hat{\bm{\pi}}$ of conditional probabilities of CpG site methylation status given gene cluster membership, (c) conditional probabilities of genes belonging to cluster $k$ given a single CpG site associated with the gene belongs to cluster $l$.}
\end{figure}
 \begin{figure}[htb]
    \centering
    \includegraphics[width=\textwidth]{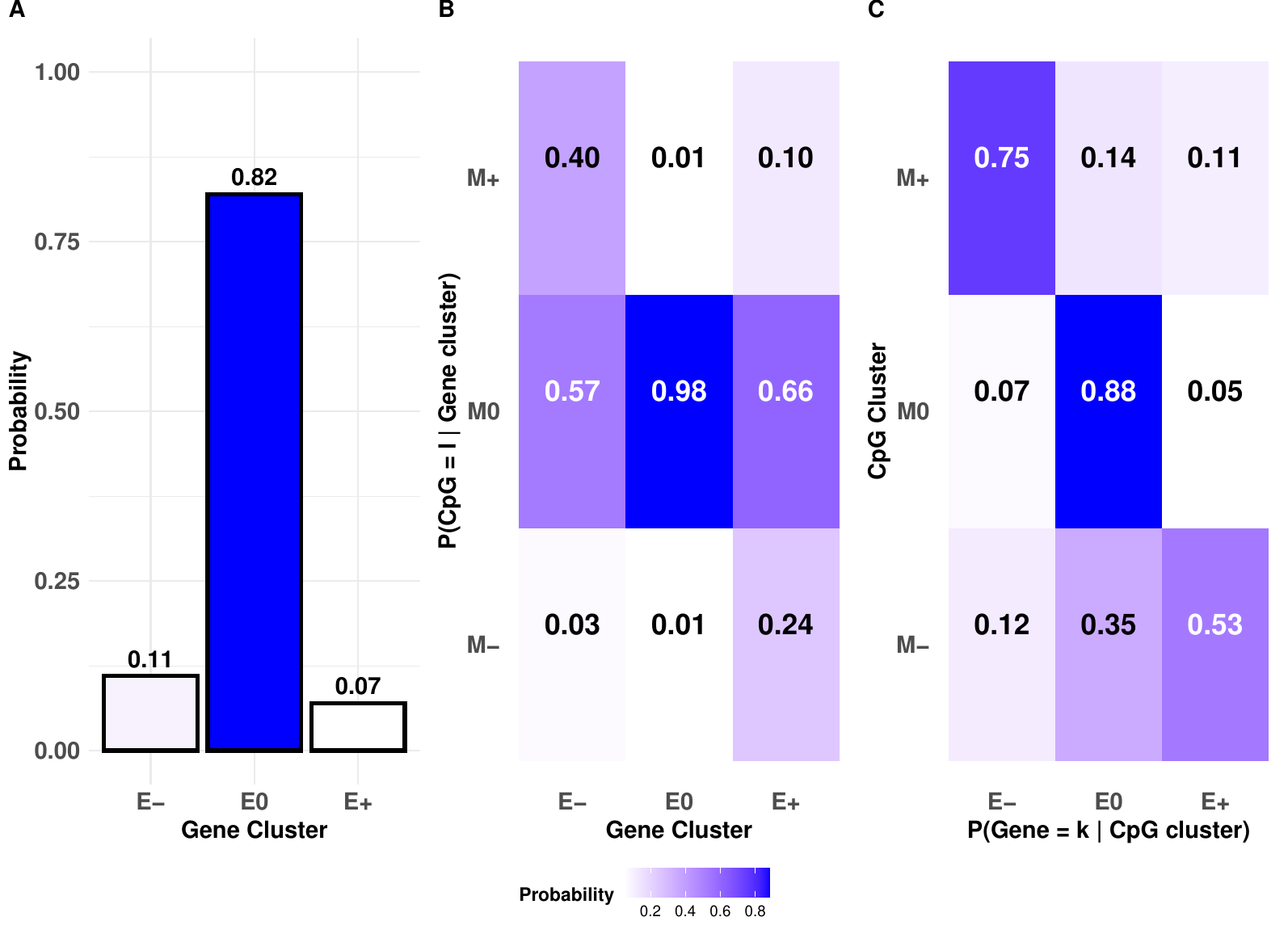}
    \caption{\texttt{Idiffomix} applied to chromosome 6 of TCGA breast cancer data: (A) estimated cluster  membership probabilities $\hat{\bm{\tau}}$, (B) the estimated matrix $\hat{\bm{\pi}}$ of conditional probabilities of CpG site methylation status given gene cluster membership, (c) conditional probabilities of genes belonging to cluster $k$ given a single CpG site associated with the gene belongs to cluster $l$.}
\end{figure}
 \begin{figure}[htb]
    \centering
    \includegraphics[width=\textwidth]{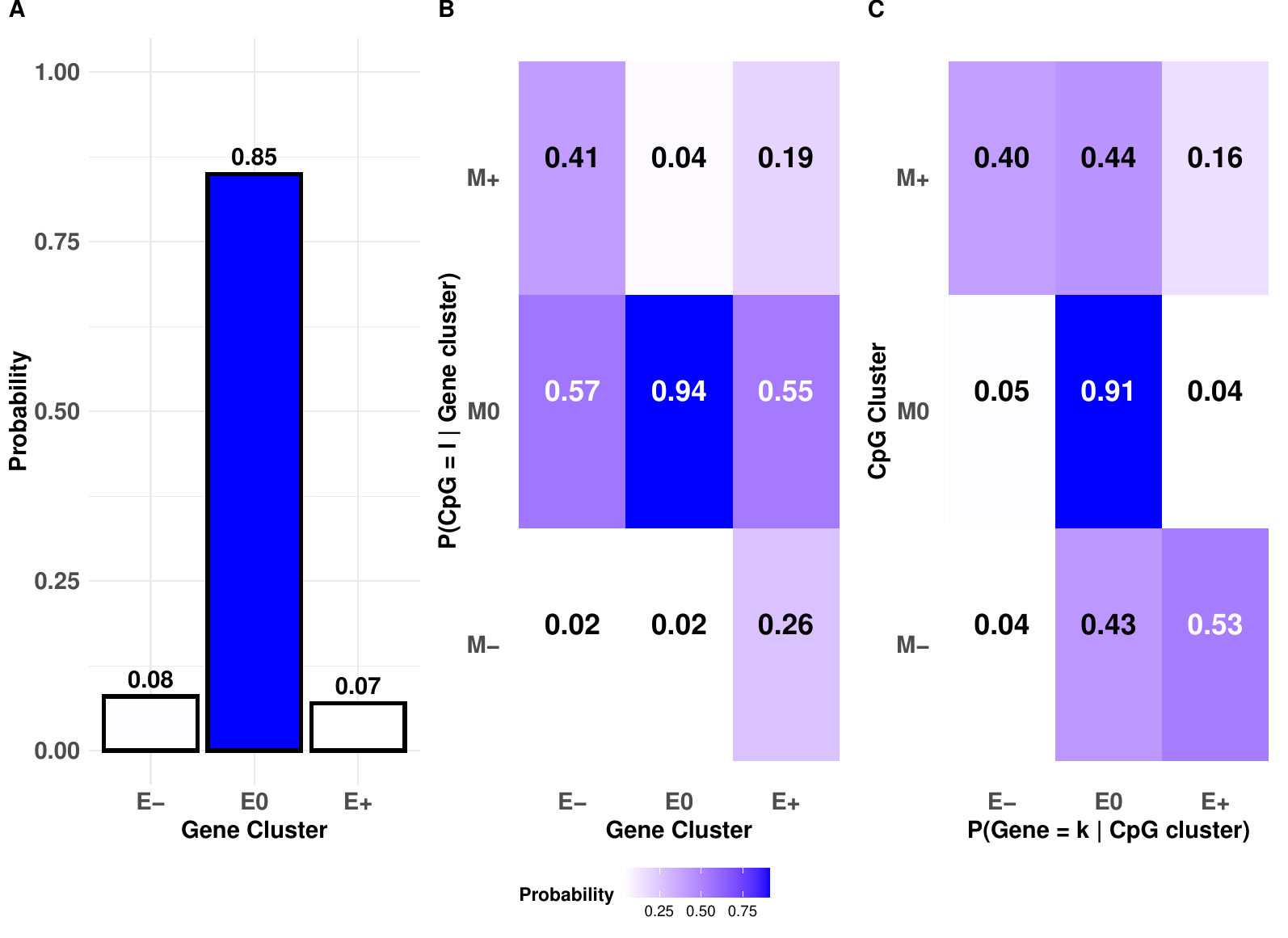}
    \caption{\texttt{Idiffomix} applied to chromosome 8 of TCGA breast cancer data: (A) estimated cluster  membership probabilities $\hat{\bm{\tau}}$, (B) the estimated matrix $\hat{\bm{\pi}}$ of conditional probabilities of CpG site methylation status given gene cluster membership, (c) conditional probabilities of genes belonging to cluster $k$ given a single CpG site associated with the gene belongs to cluster $l$.}
\end{figure}
 \begin{figure}[htb]
    \centering
    \includegraphics[width=\textwidth]{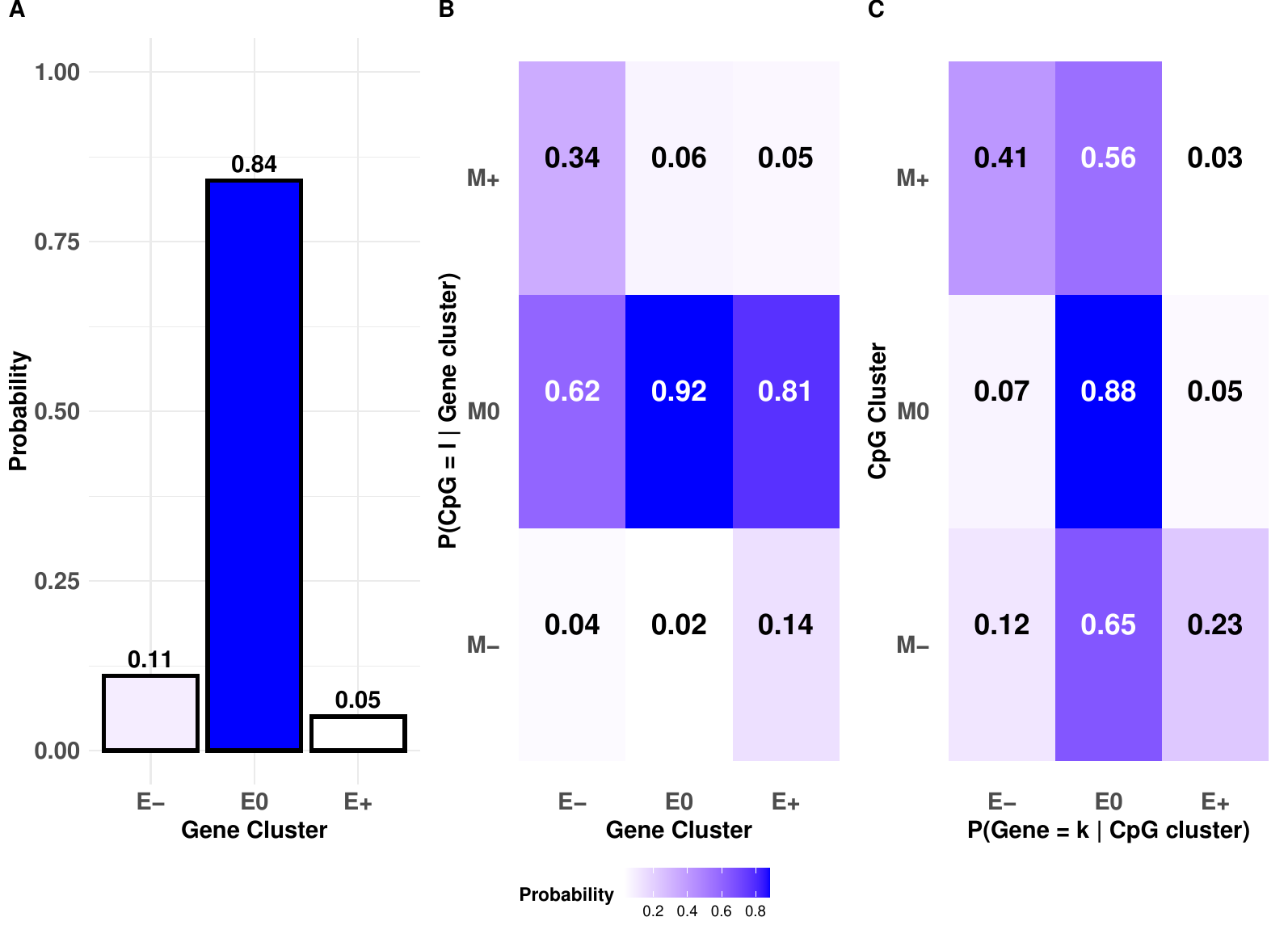}
    \caption{\texttt{Idiffomix} applied to chromosome 9 of TCGA breast cancer data: (A) estimated cluster  membership probabilities $\hat{\bm{\tau}}$, (B) the estimated matrix $\hat{\bm{\pi}}$ of conditional probabilities of CpG site methylation status given gene cluster membership, (c) conditional probabilities of genes belonging to cluster $k$ given a single CpG site associated with the gene belongs to cluster $l$.}
\end{figure}
 \begin{figure}[htb]
    \centering
    \includegraphics[width=\textwidth]{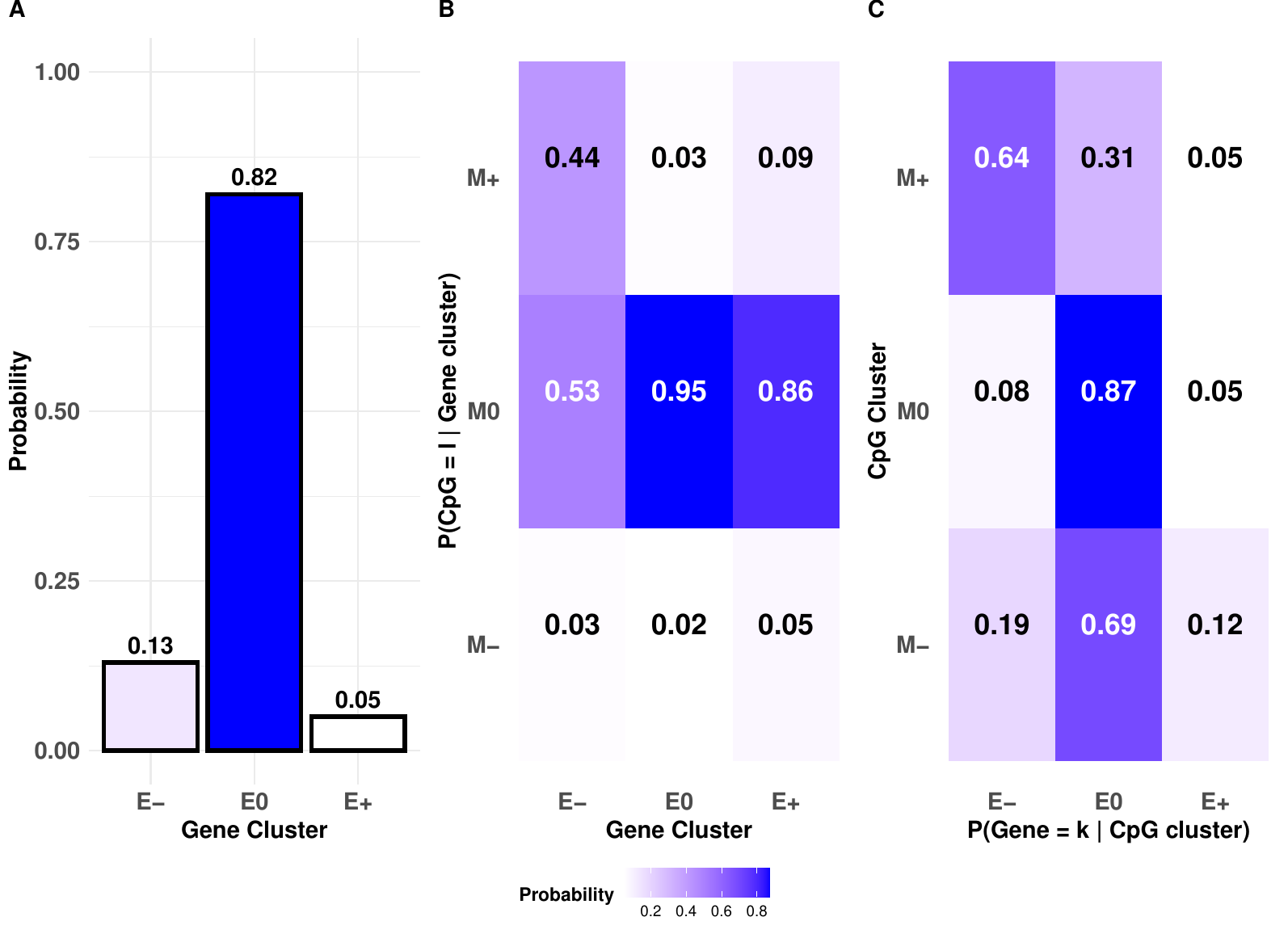}
    \caption{\texttt{Idiffomix} applied to chromosome 10 of TCGA breast cancer data: (A) estimated cluster  membership probabilities $\hat{\bm{\tau}}$, (B) the estimated matrix $\hat{\bm{\pi}}$ of conditional probabilities of CpG site methylation status given gene cluster membership, (c) conditional probabilities of genes belonging to cluster $k$ given a single CpG site associated with the gene belongs to cluster $l$.}
\end{figure}
 \begin{figure}[htb]
    \centering
    \includegraphics[width=\textwidth]{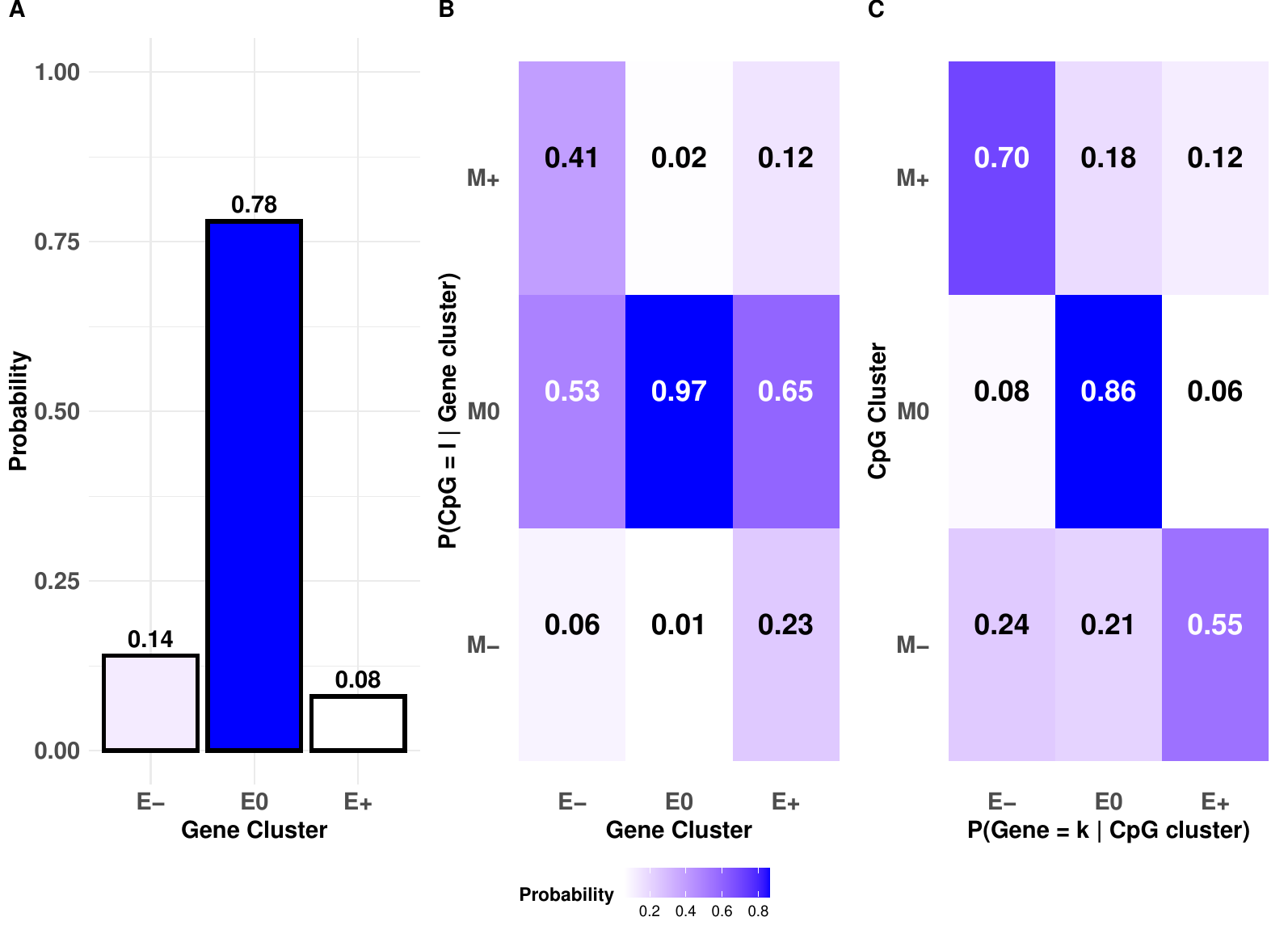}
    \caption{\texttt{Idiffomix} applied to chromosome 11 of TCGA breast cancer data: (A) estimated cluster  membership probabilities $\hat{\bm{\tau}}$, (B) the estimated matrix $\hat{\bm{\pi}}$ of conditional probabilities of CpG site methylation status given gene cluster membership, (c) conditional probabilities of genes belonging to cluster $k$ given a single CpG site associated with the gene belongs to cluster $l$.}
\end{figure}
 \begin{figure}[htb]
    \centering
    \includegraphics[width=\textwidth]{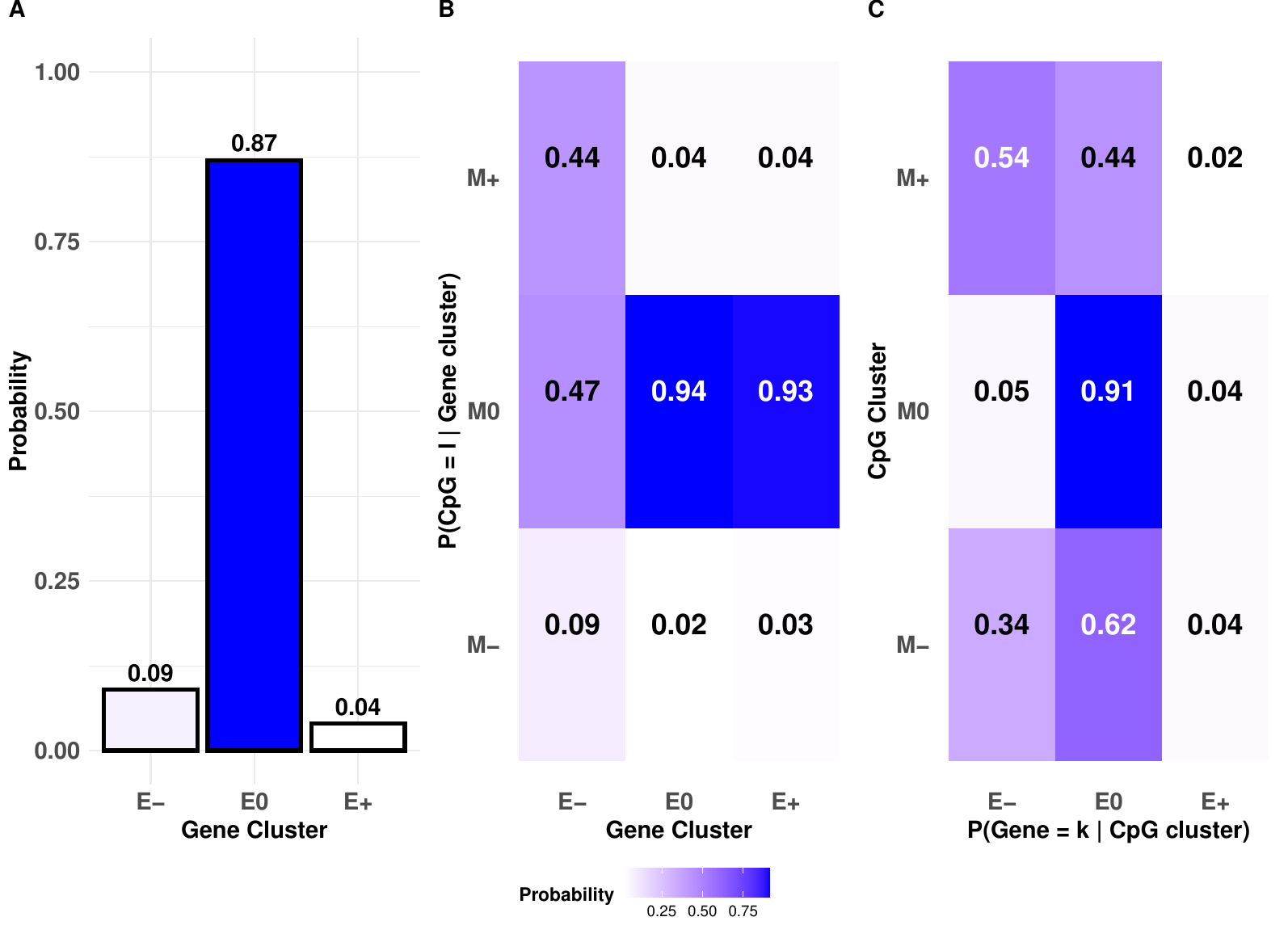}
    \caption{\texttt{Idiffomix} applied to chromosome 12 of TCGA breast cancer data: (A) estimated cluster  membership probabilities $\hat{\bm{\tau}}$, (B) the estimated matrix $\hat{\bm{\pi}}$ of conditional probabilities of CpG site methylation status given gene cluster membership, (c) conditional probabilities of genes belonging to cluster $k$ given a single CpG site associated with the gene belongs to cluster $l$.}
\end{figure}
 \begin{figure}[htb]
    \centering
    \includegraphics[width=\textwidth]{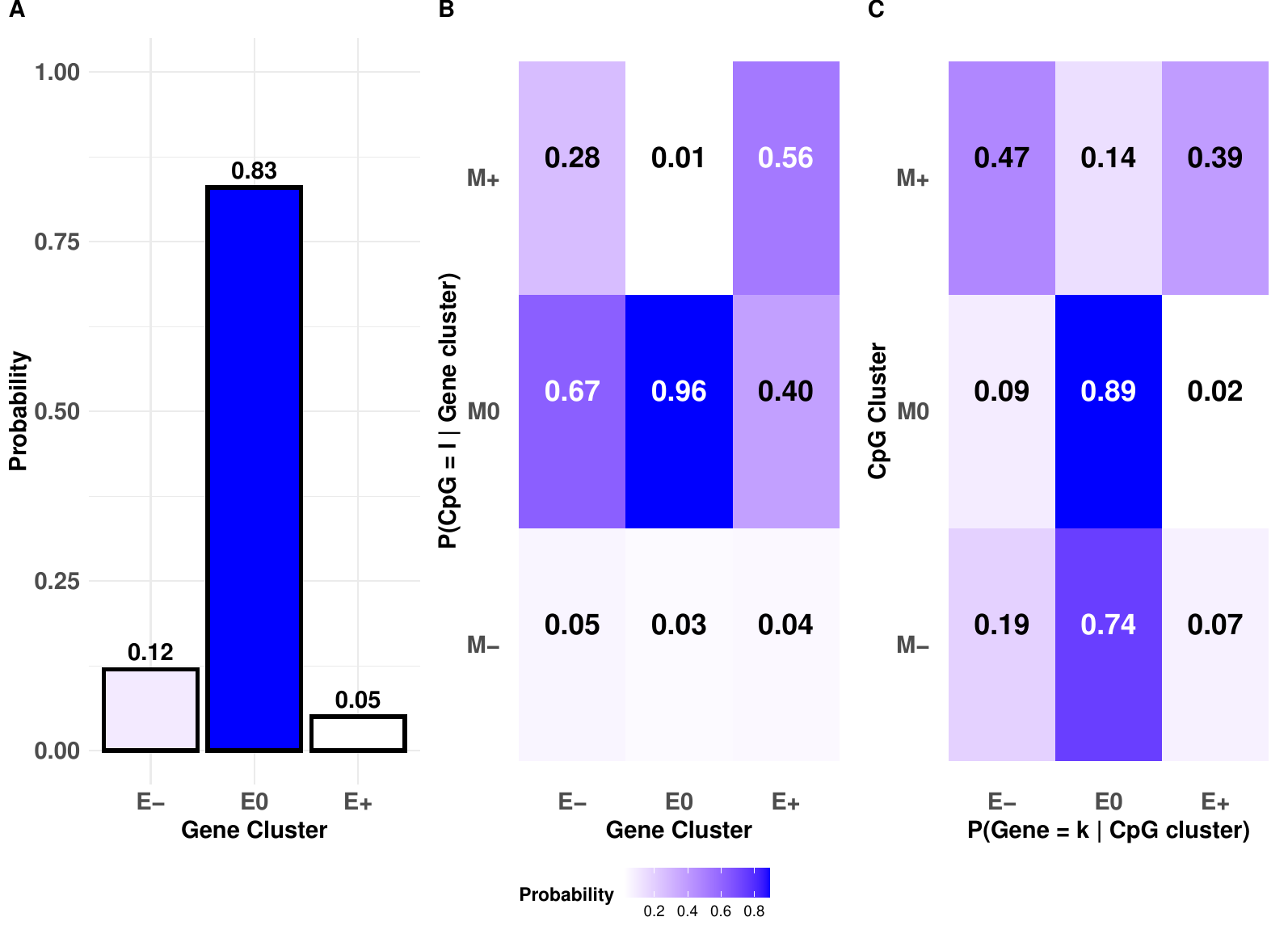}
    \caption{\texttt{Idiffomix} applied to chromosome 13 of TCGA breast cancer data: (A) estimated cluster  membership probabilities $\hat{\bm{\tau}}$, (B) the estimated matrix $\hat{\bm{\pi}}$ of conditional probabilities of CpG site methylation status given gene cluster membership, (c) conditional probabilities of genes belonging to cluster $k$ given a single CpG site associated with the gene belongs to cluster $l$.}
\end{figure}
 \begin{figure}[htb]
    \centering
    \includegraphics[width=\textwidth]{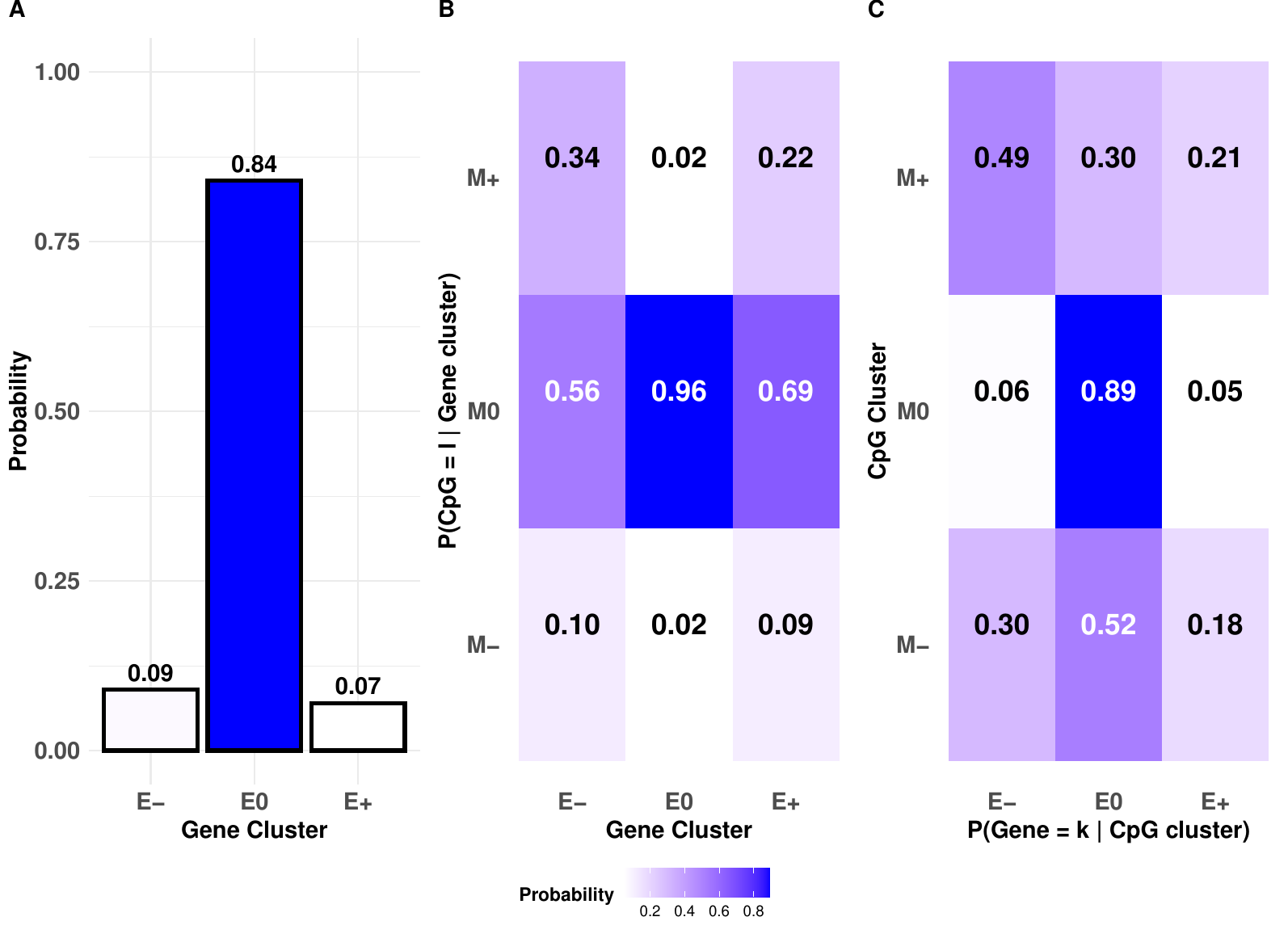}
    \caption{\texttt{Idiffomix} applied to chromosome 14 of TCGA breast cancer data: (A) estimated cluster  membership probabilities $\hat{\bm{\tau}}$, (B) the estimated matrix $\hat{\bm{\pi}}$ of conditional probabilities of CpG site methylation status given gene cluster membership, (c) conditional probabilities of genes belonging to cluster $k$ given a single CpG site associated with the gene belongs to cluster $l$.}
\end{figure} 

\begin{figure}[htb]
    \centering
    \includegraphics[width=\textwidth]{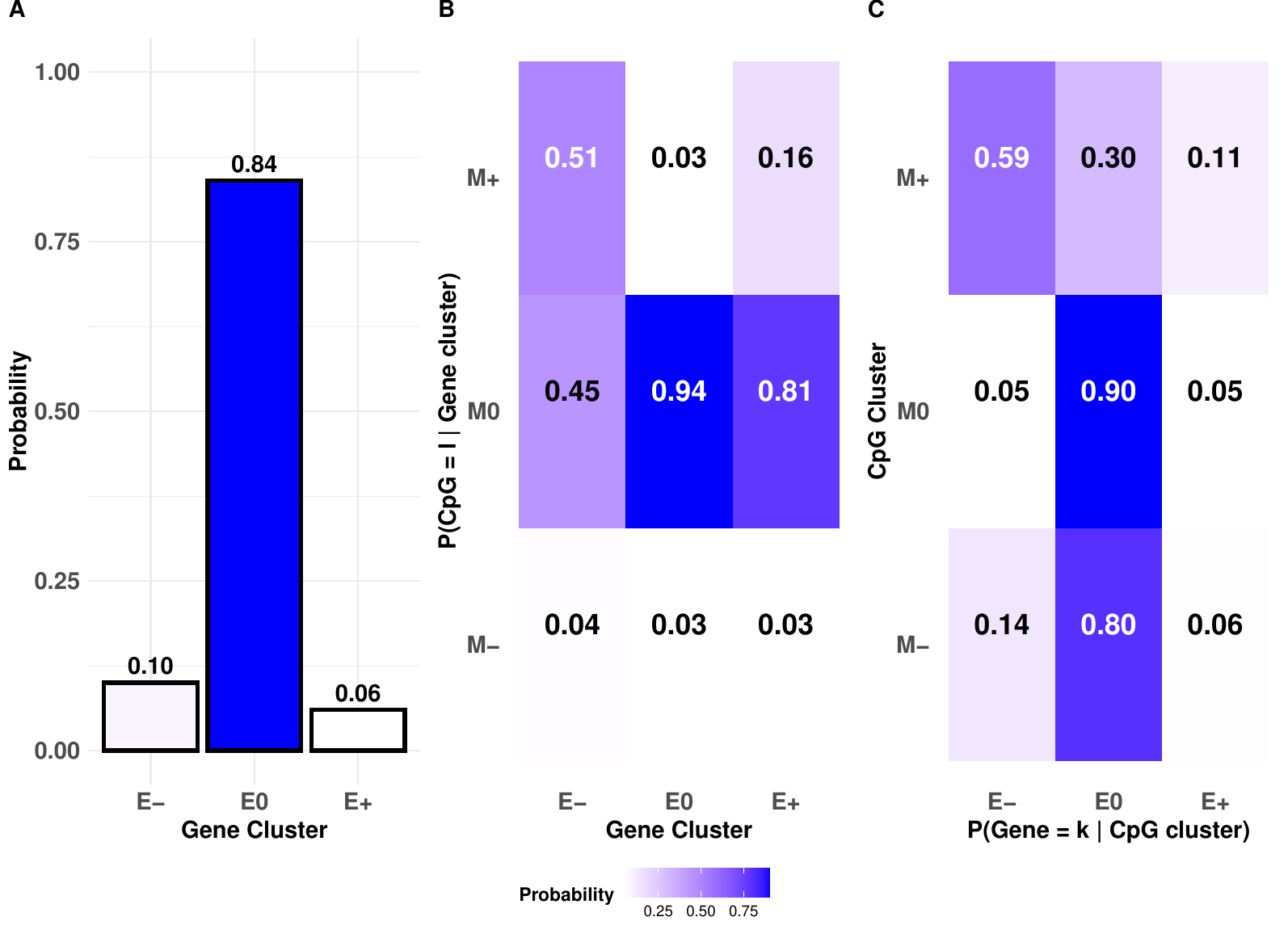}
    \caption{\texttt{Idiffomix} applied to chromosome 15 of TCGA breast cancer data: (A) estimated cluster  membership probabilities $\hat{\bm{\tau}}$, (B) the estimated matrix $\hat{\bm{\pi}}$ of conditional probabilities of CpG site methylation status given gene cluster membership, (c) conditional probabilities of genes belonging to cluster $k$ given a single CpG site associated with the gene belongs to cluster $l$.}
\end{figure} 

\begin{figure}[htb]
    \centering
    \includegraphics[width=\textwidth]{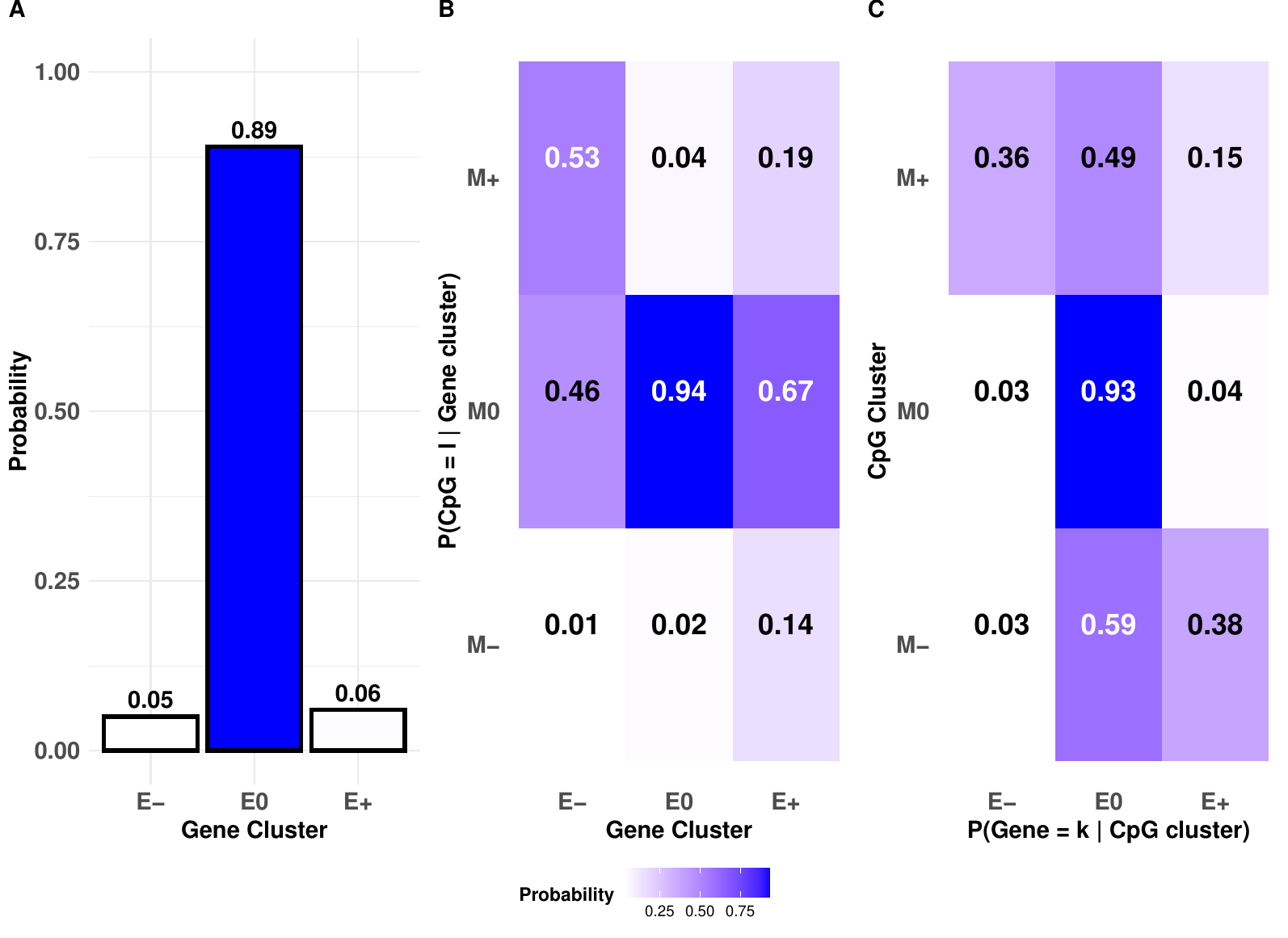}
    \caption{\texttt{Idiffomix} applied to chromosome 16 of TCGA breast cancer data: (A) estimated cluster  membership probabilities $\hat{\bm{\tau}}$, (B) the estimated matrix $\hat{\bm{\pi}}$ of conditional probabilities of CpG site methylation status given gene cluster membership, (c) conditional probabilities of genes belonging to cluster $k$ given a single CpG site associated with the gene belongs to cluster $l$.}
\end{figure}

 \begin{figure}[htb]
    \centering
    \includegraphics[width=\textwidth]{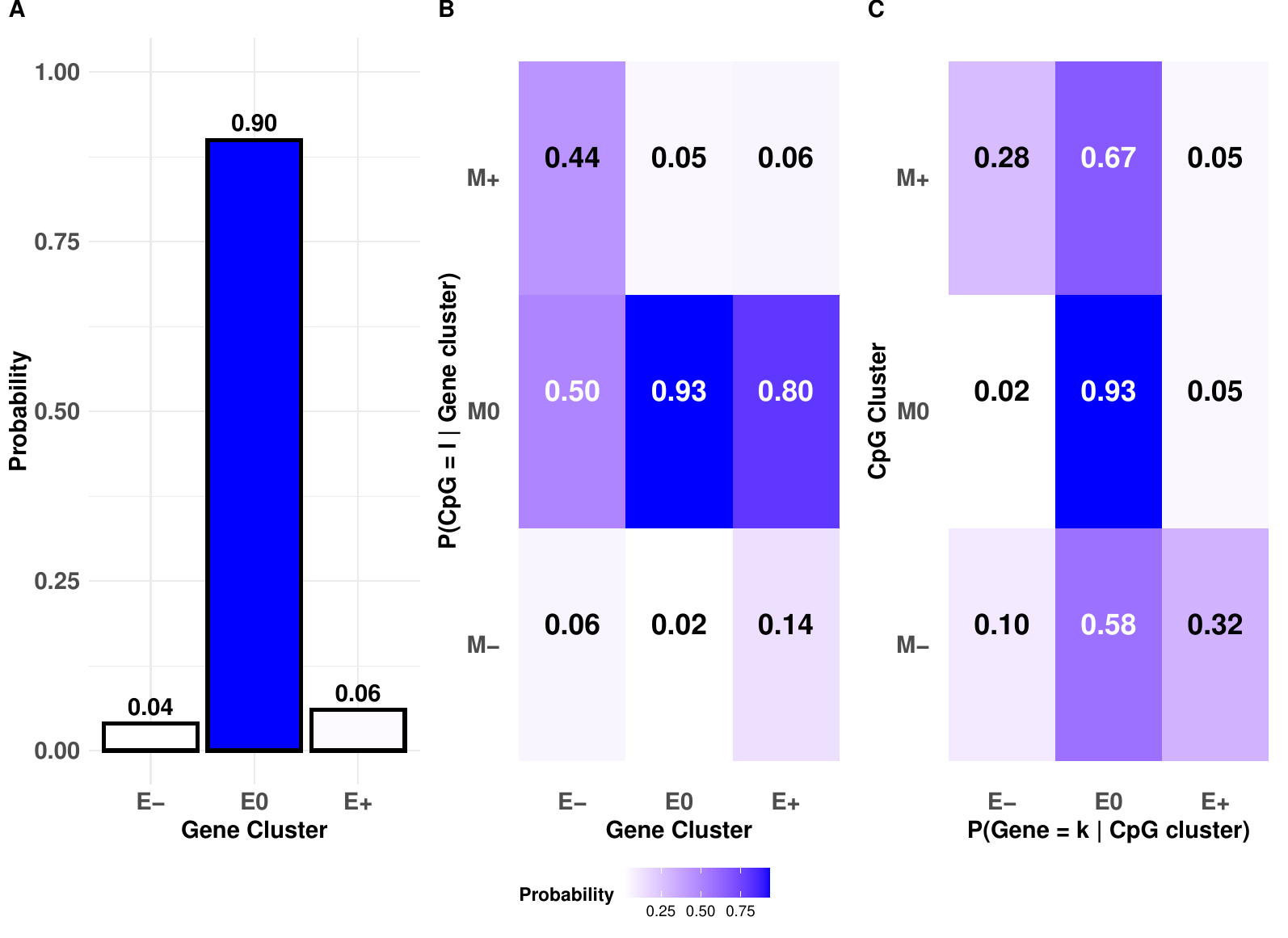}
    \caption{\texttt{Idiffomix} applied to chromosome 17 of TCGA breast cancer data: (A) estimated cluster  membership probabilities $\hat{\bm{\tau}}$, (B) the estimated matrix $\hat{\bm{\pi}}$ of conditional probabilities of CpG site methylation status given gene cluster membership, (c) conditional probabilities of genes belonging to cluster $k$ given a single CpG site associated with the gene belongs to cluster $l$.}
\end{figure}
 \begin{figure}[htb]
    \centering
    \includegraphics[width=\textwidth]{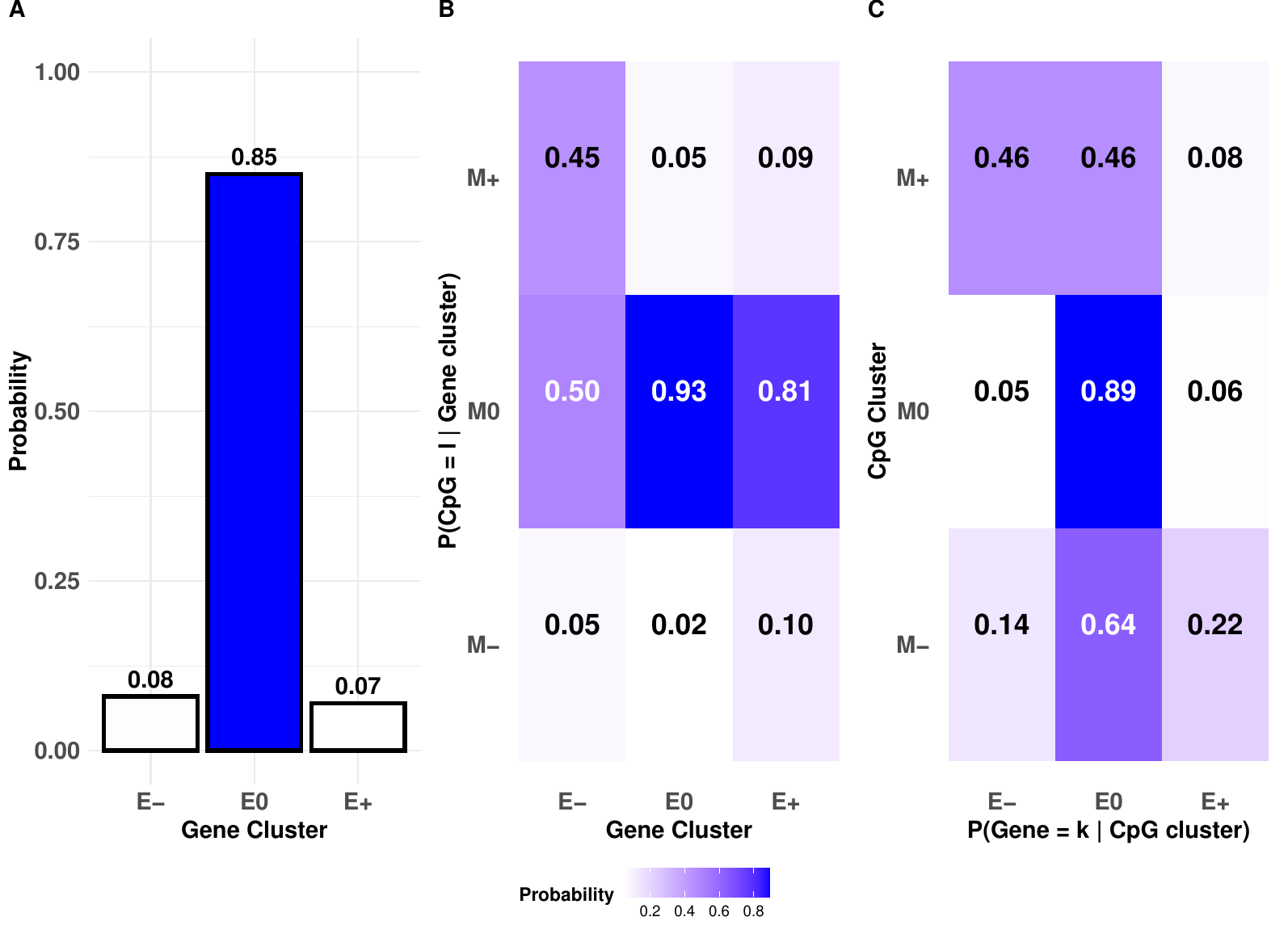}
    \caption{\texttt{Idiffomix} applied to chromosome 18 of TCGA breast cancer data: (A) estimated cluster  membership probabilities $\hat{\bm{\tau}}$, (B) the estimated matrix $\hat{\bm{\pi}}$ of conditional probabilities of CpG site methylation status given gene cluster membership, (c) conditional probabilities of genes belonging to cluster $k$ given a single CpG site associated with the gene belongs to cluster $l$.}
\end{figure}
 \begin{figure}[htb]
    \centering
    \includegraphics[width=\textwidth]{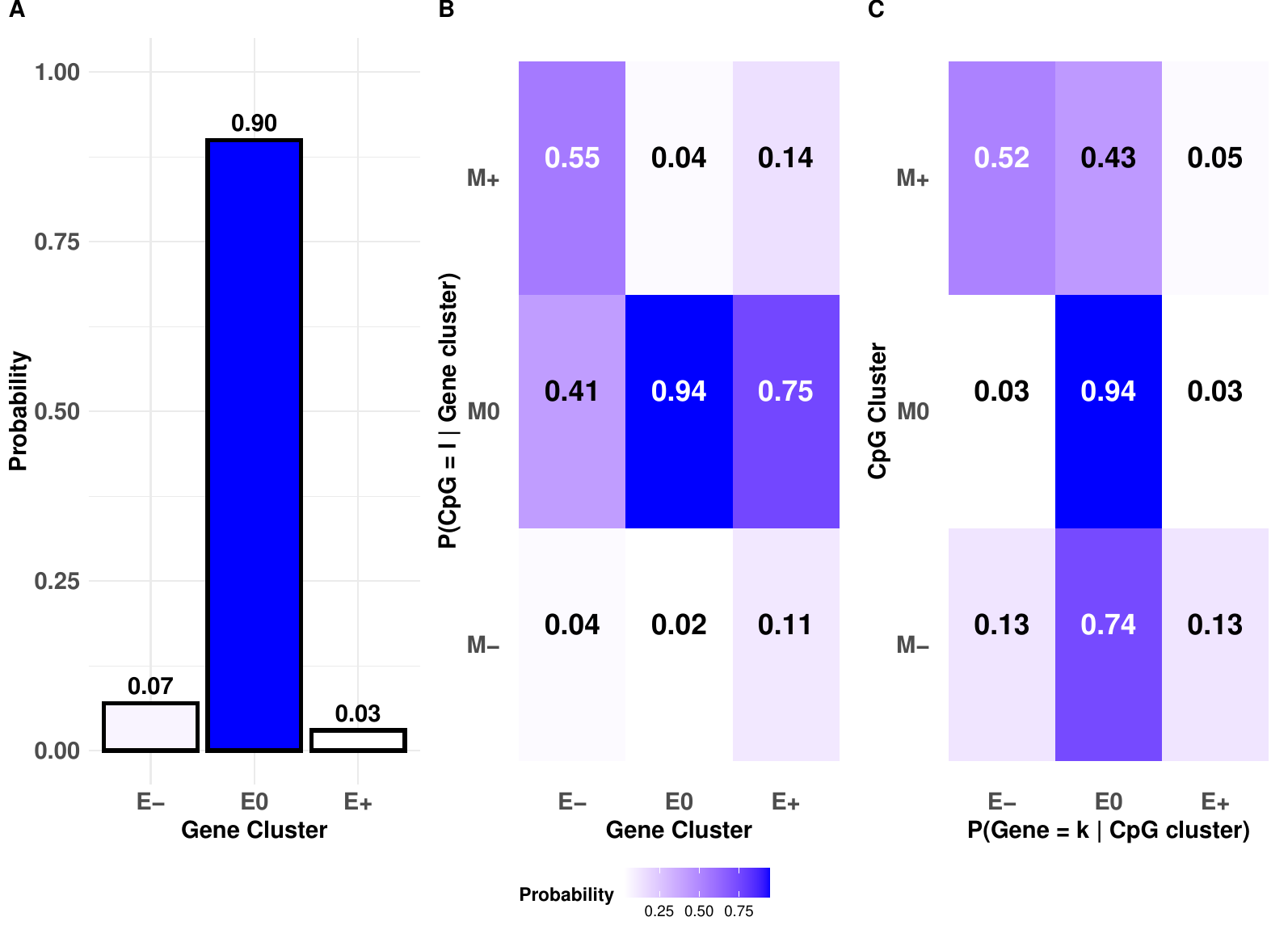}
    \caption{\texttt{Idiffomix} applied to chromosome 19 of TCGA breast cancer data: (A) estimated cluster  membership probabilities $\hat{\bm{\tau}}$, (B) the estimated matrix $\hat{\bm{\pi}}$ of conditional probabilities of CpG site methylation status given gene cluster membership, (c) conditional probabilities of genes belonging to cluster $k$ given a single CpG site associated with the gene belongs to cluster $l$.}
\end{figure}
 \begin{figure}[htb]
    \centering
    \includegraphics[width=\textwidth]{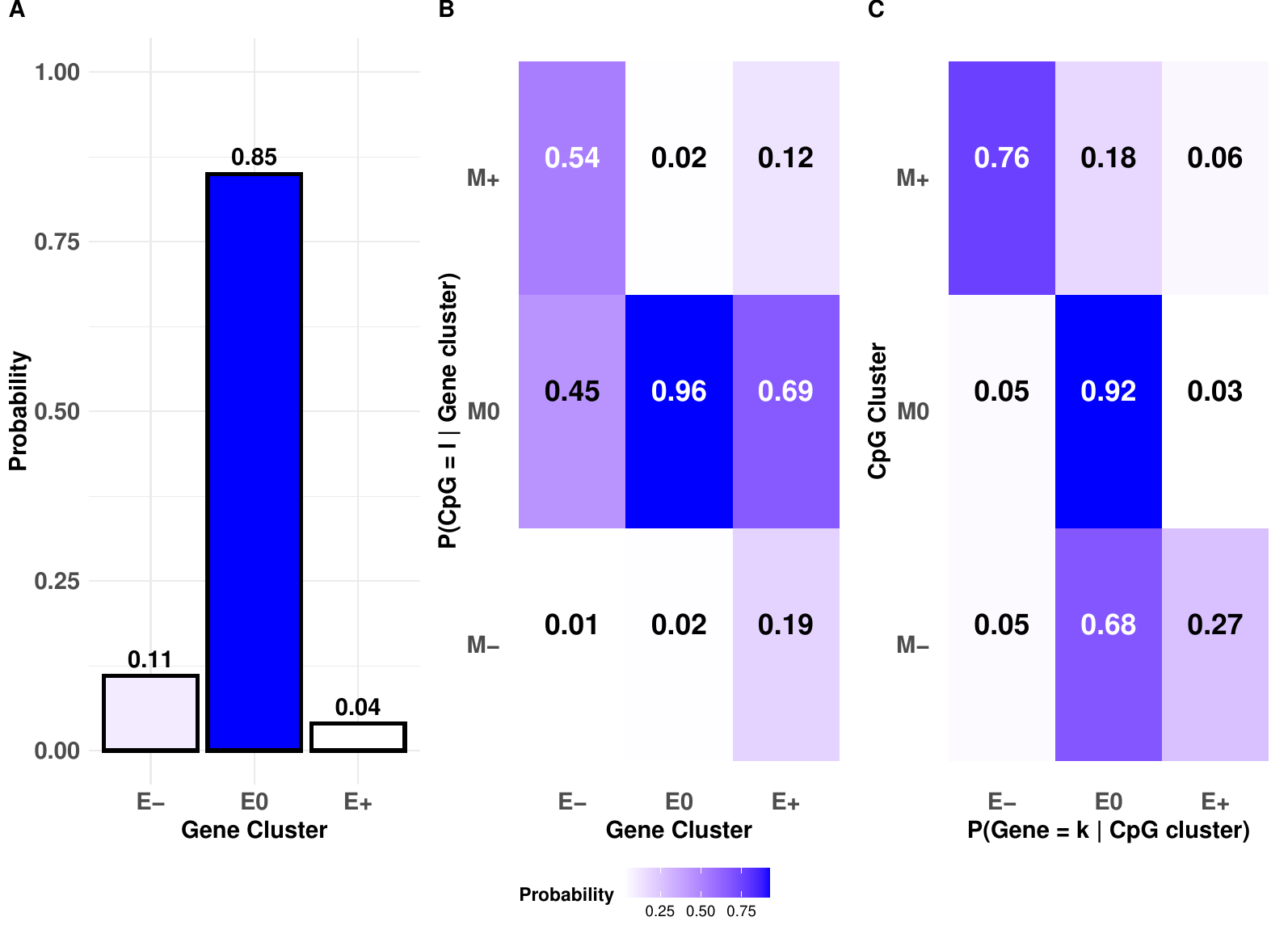}
    \caption{\texttt{Idiffomix} applied to chromosome 20 of TCGA breast cancer data: (A) estimated cluster  membership probabilities $\hat{\bm{\tau}}$, (B) the estimated matrix $\hat{\bm{\pi}}$ of conditional probabilities of CpG site methylation status given gene cluster membership, (c) conditional probabilities of genes belonging to cluster $k$ given a single CpG site associated with the gene belongs to cluster $l$.}
\end{figure}
 \begin{figure}[htb]
    \centering
    \includegraphics[width=\textwidth]{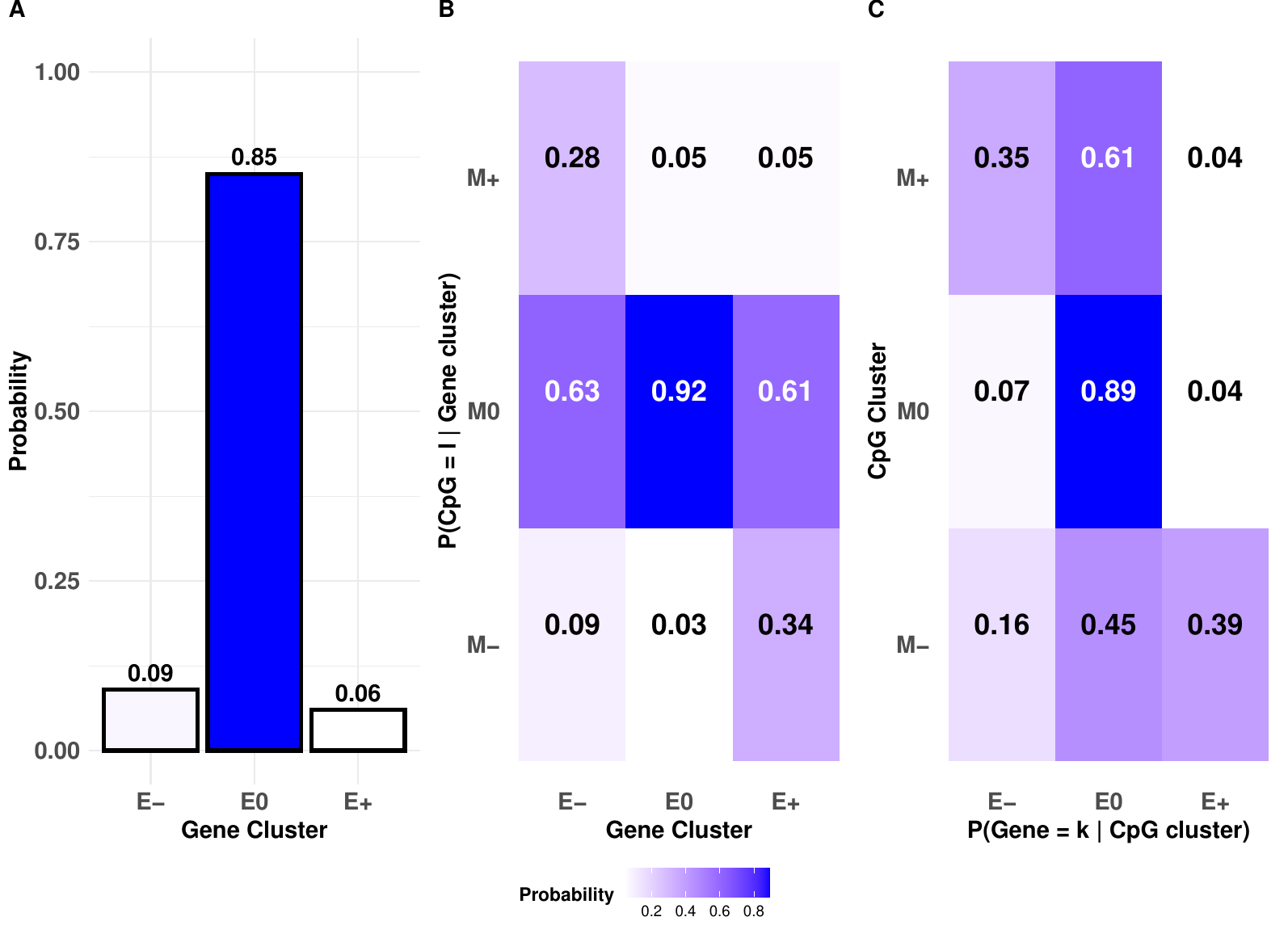}
    \caption{\texttt{Idiffomix} applied to chromosome 21 of TCGA breast cancer data: (A) estimated cluster  membership probabilities $\hat{\bm{\tau}}$, (B) the estimated matrix $\hat{\bm{\pi}}$ of conditional probabilities of CpG site methylation status given gene cluster membership, (c) conditional probabilities of genes belonging to cluster $k$ given a single CpG site associated with the gene belongs to cluster $l$.}
\end{figure}
 \begin{figure}[htb]
    \centering
    \includegraphics[width=\textwidth]{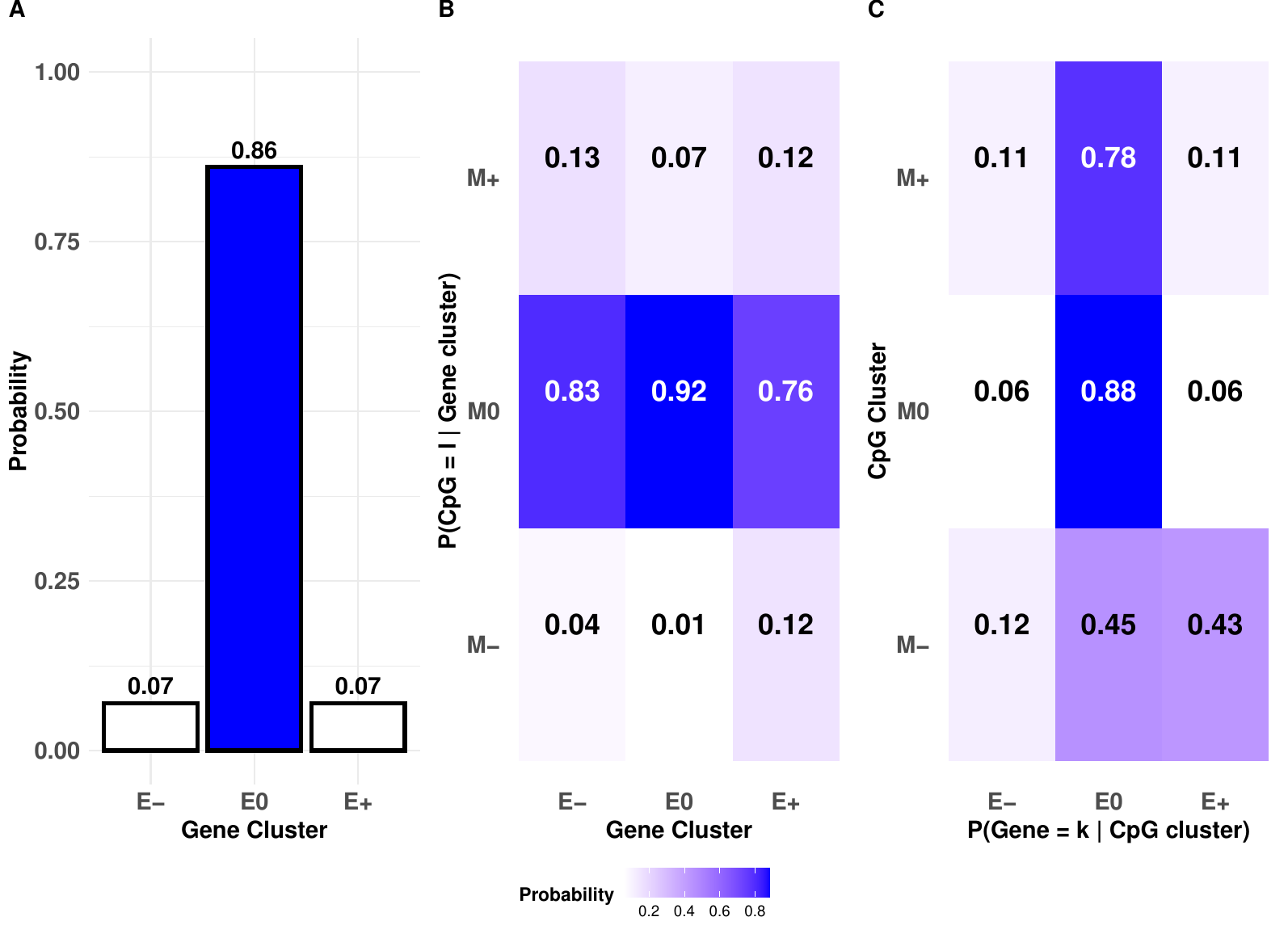}
    \caption{\texttt{Idiffomix} applied to chromosome 22 of TCGA breast cancer data: (A) estimated cluster  membership probabilities $\hat{\bm{\tau}}$, (B) the estimated matrix $\hat{\bm{\pi}}$ of conditional probabilities of CpG site methylation status given gene cluster membership, (c) conditional probabilities of genes belonging to cluster $k$ given a single CpG site associated with the gene belongs to cluster $l$.}
\end{figure}

\clearpage

\subsection{Clustering uncertainty}
 \begin{figure}[h!]
    \centering
    \includegraphics[width=\textwidth]{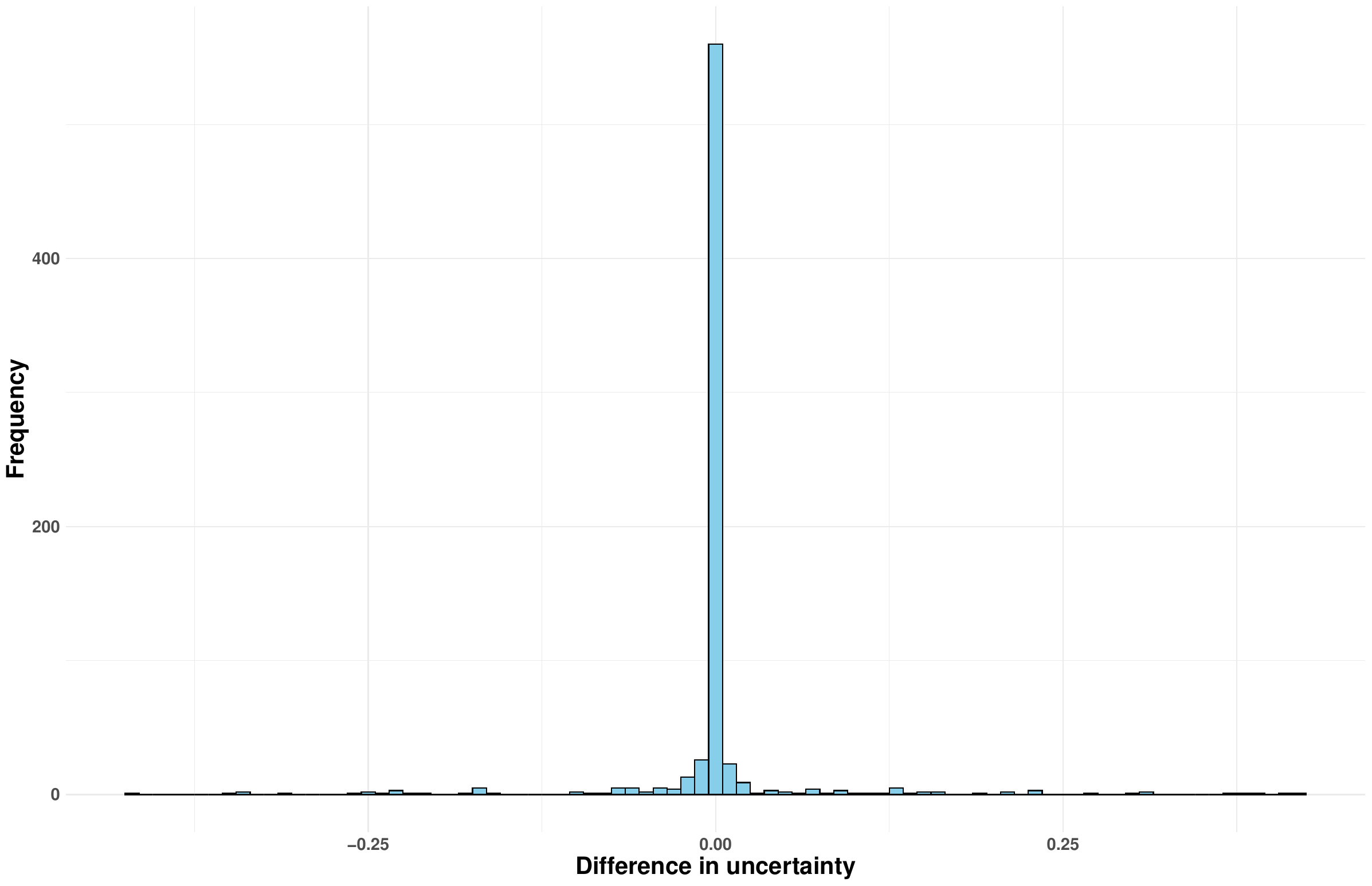}
    \caption{Difference in uncertainties for genes not changing clustering between \texttt{idiffomix} and \texttt{mclust}.}
\end{figure}

\clearpage

\subsection{Other genes of interest}

\begin{figure}[h!]
    \centering
    \includegraphics[width=\textwidth]{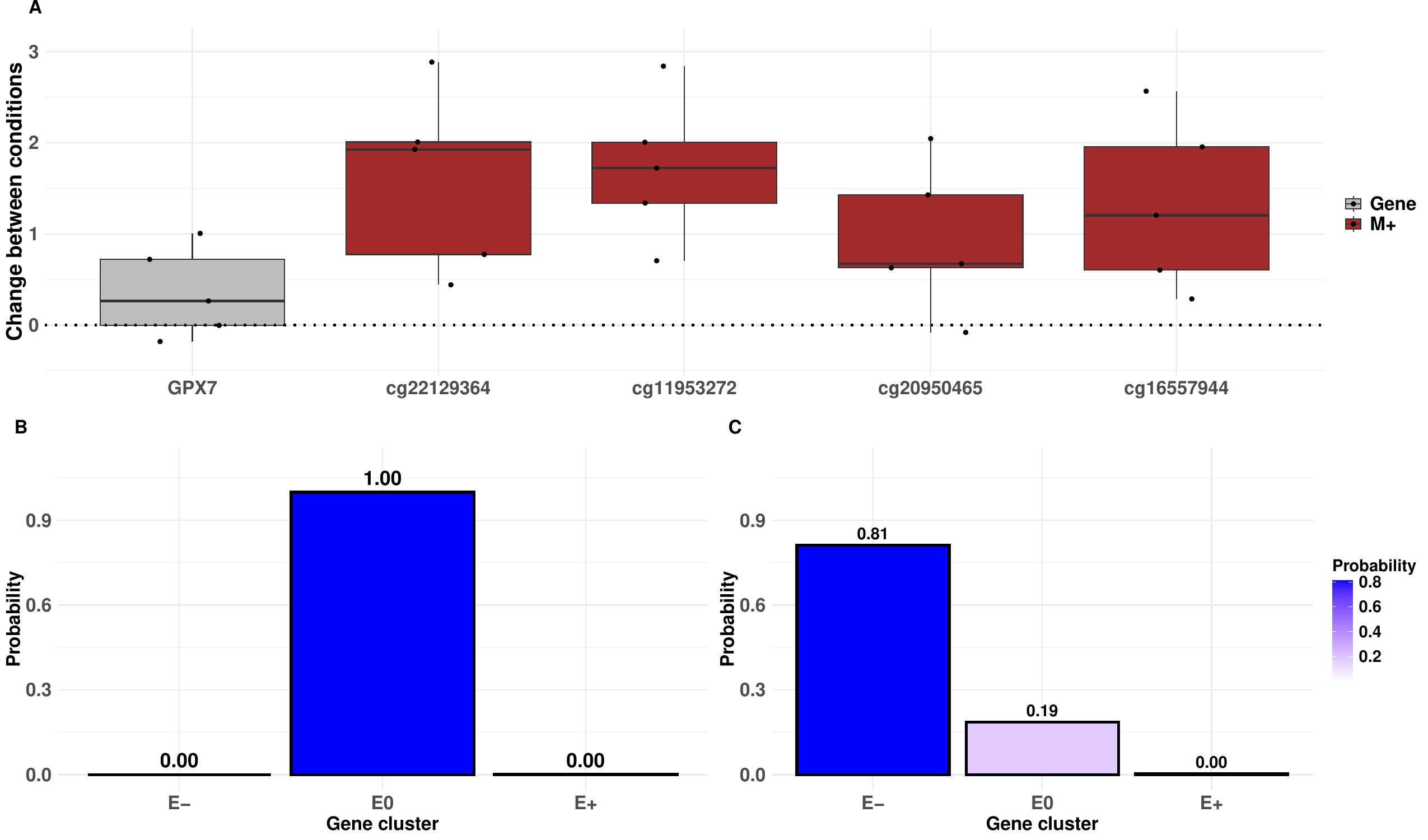}
    \caption{Comparison of results for independent and integrated analyses for GPX7 on chromosome 1: (A) log-fold change in gene expression levels (grey) and differences in \textit{M}-values between tumour and normal samples, coloured by inferred idiffomix cluster (hypermethylated CpG sites, M+ in brown); (B) posterior probability of \textit{GPX7} belonging to the E-, E0 and E+ clusters under \texttt{mclust}, (C) posterior probability of \textit{GPX7} belonging to the E-, E0 and E+ clusters when jointly modelled with methylation data under \texttt{idiffomix}. Larger posterior probabilities are represented by increasingly dark shades of blue}
    \label{idiffomix_appfig:GPX7}
\end{figure}

Genes implicated in breast cancer, such as \textit{TNFRSF18}, \textit{GPX7}, and \textit{RAD51} play important roles in the development and progression of the disease.  Panel A in Figure \ref{idiffomix_appfig:GPX7} illustrates that the log-fold change of the gene expression for \textit{GPX7} and the differences in \textit{M}-values for the associated CpG sites, located on chromosome 1. When the gene expression data is modelled independently Panel B in Figure \ref{idiffomix_appfig:GPX7} suggested \textit{GPX7} to be non-differential (E0). However, the differences in \textit{M}-values of the CpG sites linked to this gene are suggested to be hypermethylated (M+) when modelled jointly. Therefore, when the two data types are modelled jointly, Panel C in Figure \ref{idiffomix_appfig:GPX7} suggests the gene to be downregulated (E-) with a probability of 0.81 and to be non-differential (E0) with a probability of $0.19$. 

\begin{figure}[h!]
    \centering
    \includegraphics[width=\textwidth]{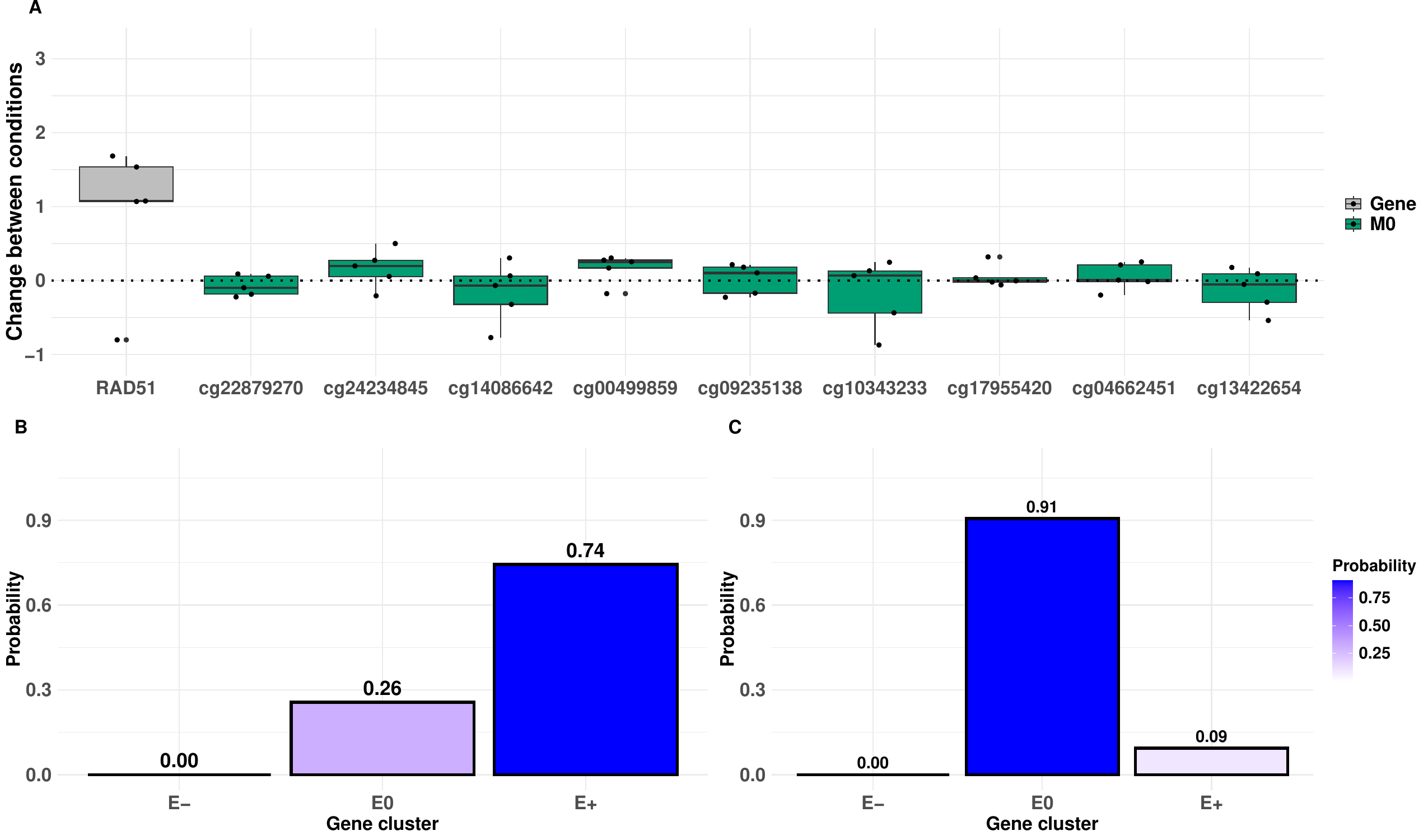}
    \caption{Comparison of results for independent and integrated analyses for \textit{RAD51} on chromosome 15: (A) log-fold change in gene expression levels (grey) and differences in \textit{M}-values between tumour and normal samples, coloured by inferred idiffomix cluster (non-differentially methylated CpG sites, M0 in green); (B) posterior probability of \textit{RAD51} belonging to the E-, E0 and E+ clusters under \texttt{mclust}, (C) posterior probability of \textit{RAD51} belonging to the E-, E0 and E+ clusters when jointly modelled with methylation data under \texttt{idiffomix}. Larger posterior probabilities are represented by increasingly dark shades of blue.}
    \label{idiffomix_appfig:RAD51}
\end{figure}

Panel A in Figure \ref{idiffomix_appfig:RAD51} illustrates the log-fold change for RAD51 gene, located on chromosome 15, and the differences in \textit{M}-values for the associated CpG sites. Panel B in Figure \ref{idiffomix_appfig:RAD51} suggests the gene to be upregulated when modelled independently and is estimated to be in cluster E+ with probability of 0.74. However, the CpG sites linked to this gene are all found to be non-differential (M0) when the two data types are modelled jointly. Thus, Panel C in Figure \ref{idiffomix_appfig:RAD51} suggests the gene to be in E0 cluster with a probability of 0.9 and in E+ cluster with a probability of 0.1 when the gene expression and methylation data are modelled jointly.

\clearpage

\subsection{Top GO and KEGG terms associated with identified DEGs and DMCs}
\begin{table}[htbp]
\centering
\caption{Top 10 GO processes linked to DMCs identified by \texttt{idiffomix} and not \texttt{limma}.}
\begin{tabularx}{\textwidth}{llXl}
\hline
\textbf{GO Process} & \textbf{ONTOLOGY} & \textbf{TERM} & \textbf{FDR} \\ \hline
GO:0062023 & CC & collagen-containing extracellular matrix & $4.91\times 10^{-13}$ \\ \hline
GO:0051094 & BP & positive regulation of developmental process & $2.68 \times 10^{-11}$ \\ \hline
GO:0045229 & BP & external encapsulating structure organization & $3.28 \times 10^{-11}$ \\ \hline
GO:0007166 & BP & cell surface receptor signaling pathway & $4.56\times 10^{-11}$ \\ \hline
GO:0030198 & BP & extracellular matrix organization & $8.04\times 10^{-11}$ \\ \hline
GO:0043062 & BP & extracellular structure organization & $1.31\times 10^{-10}$ \\ \hline
GO:0048513 & BP & animal organ development & $3.65\times 10^{-10}$ \\ \hline
GO:0048646 & BP & anatomical structure formation involved in morphogenesis & $5.432\times 10^{-10}$ \\ \hline
GO:0005201 & MF & extracellular matrix structural constituent & $1.07\times 10^{-9}$ \\ \hline
GO:0030545 & MF & signaling receptor regulator activity & $1.5\times 10^{-09}$ \\ \hline
\end{tabularx}

\end{table}

\begin{table}[htbp]
\centering
\caption{Top 10 GO processes linked to DMCs identified by \texttt{idiffomix} and not \texttt{mclust}.}
\begin{tabularx}{\textwidth}{llXl}
\hline
\textbf{GO Process} & \textbf{ONTOLOGY} & \textbf{TERM} & \textbf{FDR} \\ \hline
GO:0030509 & BP & BMP signaling pathway & 0.0023 \\ \hline
GO:0018146 & BP & keratan sulfate biosynthetic process & 0.0040 \\ \hline
GO:0042339 & BP & keratan sulfate metabolic process & 0.0042 \\ \hline
GO:0035282 & BP & segmentation & 0.0056 \\ \hline
GO:0010720 & BP & positive regulation of cell development & 0.0060 \\ \hline
GO:0043950 & BP & positive regulation of cAMP-mediated signaling & 0.0081 \\ \hline
GO:0000165 & BP & MAPK cascade & 0.0082 \\ \hline
GO:0070371 & BP & ERK1 and ERK2 cascade & 0.0128 \\ \hline
GO:0001707 & BP & mesoderm formation & 0.0134 \\ \hline
GO:0035418 & BP & protein localization to synapse & 0.0144 \\ \hline
\end{tabularx}

\end{table}

\begin{table}[htbp]
\centering
\caption{Top 10 KEGG pathways linked to DMCs identified by \texttt{idiffomix} and not \texttt{limma}.}
\begin{tabularx}{\textwidth}{lXl}
\hline
\textbf{KEGG Pathway} & \textbf{TERM} & \textbf{FDR} \\ \hline
hsa04820 & Cytoskeleton in muscle cells & $1.41\times 10^{-05}$ \\ \hline
hsa04060 & Cytokine-cytokine receptor interaction & $3.42\times 10^{-05}$ \\ \hline
hsa05033 & Nicotine addiction & $3.42\times 10^{-05}$ \\ \hline
hsa04974 & Protein digestion and absorption & $2.55\times 10^{-03}$ \\ \hline
hsa04512 & ECM-receptor interaction & $3.44\times 10^{-03}$ \\ \hline
hsa04514 & Cell adhesion molecules & $1.24\times 10^{-02}$ \\ \hline
hsa05146 & Amoebiasis & $1.53\times 10^{-02}$ \\ \hline
hsa00512 & Mucin type O-glycan biosynthesis & $1.89\times 10^{-02}$ \\ \hline
hsa04061 & Viral protein interaction with cytokine and cytokine receptor & $2.09\times 10^{-02}$ \\ \hline
hsa05032 & Morphine addiction & $2.09\times 10^{-02}$ \\ \hline
\end{tabularx}

\end{table}

\begin{table}[htbp]
\centering
\caption{Top KEGG pathways linked to DMCs identified by \texttt{idiffomix} and not \texttt{mclust}.}
\begin{tabularx}{\textwidth}{lXl}
\hline
\textbf{KEGG Pathway} & \textbf{TERM} & \textbf{FDR} \\ \hline
hsa04640  & Hematopoietic cell lineage & $0.027$ \\ \hline
hsa04724 & Glutamatergic synapse & $0.035$ \\ \hline

\end{tabularx}

\end{table}

\begin{table}[htbp]
\centering
\caption{Top 10 GO processes associated with DEGs identified by \texttt{idiffomix} and not by \texttt{limma}.}
\begin{tabularx}{\textwidth}{lXl}
\hline
\textbf{GO Process} & \textbf{TERM} & \textbf{Adjusted p-value} \\ \hline
GO:0046942 & carboxylic acid transport & $9.60\times 10^{-08}$ \\ \hline
GO:0015849 & organic acid transport & $1.034\times 10^{-07}$ \\ \hline
GO:0015711 & organic anion transport & $2.58\times 10^{-07}$ \\ \hline
GO:0009954 & proximal/distal pattern formation & $3.32\times 10^{-07}$ \\ \hline
GO:0071805 & potassium ion transmembrane transport & $1.01\times 10^{-06}$ \\ \hline
GO:0042698 & ovulation cycle & $1.31\times 10^{-06}$ \\ \hline
GO:0098657 & import into cell & $2.42\times 10^{-06}$ \\ \hline
GO:0048645 & animal organ formation & $2.53\times 10^{-06}$ \\ \hline
GO:0071772 & response to BMP & $4.48\times 10^{-06}$ \\ \hline
GO:0071773 & cellular response to BMP stimulus & $4.48\times 10^{-06}$ \\ \hline
\end{tabularx}

\end{table}

\begin{table}[htbp]
\centering
\caption{Top 10 GO processes associated with DEGs identified by \texttt{idiffomix} and not by \texttt{mclust}.}
\begin{tabularx}{\textwidth}{lXl}
\hline
\textbf{GO Process} & \textbf{TERM} & \textbf{Adjusted p-value} \\ \hline
GO:0007215 & glutamate receptor signaling pathway & $1.66\times 10^{-04}$ \\ \hline
GO:0050808 & synapse organization & $2.58\times 10^{-04}$ \\ \hline
GO:0010092 & specification of animal organ identity & $4.13\times 10^{-04}$ \\ \hline
GO:0035136 & forelimb morphogenesis & $9.273\times 10^{-04}$ \\ \hline
GO:0021545 & cranial nerve development & $9.33\times 10^{-04}$ \\ \hline
GO:0048934 & peripheral nervous system neuron differentiation & $1.03\times 10^{-03}$ \\ \hline
GO:0048935 & peripheral nervous system neuron development & $1.03\times 10^{-03}$ \\ \hline
GO:0033555 & multicellular organismal response to stress & $1.44\times 10^{-03}$ \\ \hline
GO:0021983 & pituitary gland development & $1.52\times 10^{-03}$ \\ \hline
GO:0002920 & regulation of humoral immune response & $1.53\times 10^{-03}$ \\ \hline
\end{tabularx}

\end{table}

\begin{table}[htbp]
\centering
\caption{Top 10 KEGG processes associated with DEGs identified by \texttt{idiffomix} and not by \texttt{limma}.}
\begin{tabularx}{\textwidth}{lXl}
\hline
\textbf{KEGG Pathways} & \textbf{Description} & \textbf{Adjusted p-value} \\ \hline
hsa04974 &  Protein digestion and absorption & $1.87\times 10^{-08}$ \\ \hline
hsa04610 & Complement and coagulation cascades & $1.42\times 10^{-06}$ \\ \hline
hsa03320 & PPAR signaling pathway & $8.51\times 10^{-05}$ \\ \hline
hsa05144 & Malaria & $1.14\times 10^{-04}$ \\ \hline
hsa04390 & Hippo signaling pathway & $1.36\times 10^{-03}$ \\ \hline
hsa04724 & Glutamatergic synapse & $2.25\times 10^{-03}$ \\ \hline
hsa04061 & Viral protein interaction with cytokine and cytokine receptor & $2.77\times 10^{-03}$ \\ \hline
hsa04024 & cAMP signaling pathway & $6.29\times 10^{-03}$ \\ \hline
hsa05323 &  Rheumatoid arthritis & $1.41\times 10^{-02}$ \\ \hline
hsa05033 &  Nicotine addiction  & $1.48\times 10^{-02}$ \\ \hline
\end{tabularx}

\end{table}

\begin{table}[htbp]
\centering
\caption{Top KEGG processes associated with DEGs identified by \texttt{idiffomix} and not by \texttt{mclust}.}
\begin{tabularx}{\textwidth}{lXl}
\hline
\textbf{KEGG Pathways} & \textbf{Description} & \textbf{Adjusted p-value} \\ \hline
hsa04724 &   Glutamatergic synapse & $2.25\times 10^{-03}$ \\ \hline
hsa05033 & Nicotine addiction & $1.49\times 10^{-02}$ \\ \hline
hsa04080 &  Neuroactive ligand-receptor interaction & $1.91\times 10^{-02}$ \\ \hline
hsa02010 & ABC transporters  & $2.91\times 10^{-02}$ \\ \hline
hsa04713 & Circadian entrainment & $3.35\times 10^{-02}$ \\ \hline
hsa05412 & Arrhythmogenic right ventricular cardiomyopathy & $3.54\times 10^{-02}$ \\ \hline
\end{tabularx}

\end{table}

\clearpage




\newpage

\nocite{*}


\end{document}